\documentclass[twocolumn,trackchanges]{aastex701}
\usepackage{url}
\usepackage{multirow}

\begin{document}

\title{Mapping the Inner Milky Way with Infrared-Derived Distances to AGB Stars}

\author[0009-0007-8229-3036]{Rajorshi Bhattacharya} 
\altaffiliation{Reber Fellow at the National Radio Astronomy Observatory in Socorro.}
\affiliation{Department of Physics and Astronomy, University of New Mexico, 87131 Albuquerque, NM, USA} 
\email[show]{rbhattacharya1995@unm.edu}
\correspondingauthor{Rajorshi Bhattacharya, Subhajit Kar}

\author[0000-0001-7874-0218]{Subhajit Kar} 
\affiliation{Inter-University Centre for Astronomy and Astrophysics (IUCAA), Post Bag 4, Ganeshkhind, Pune 411007, India} 
\email[show]{subhajit.kar@iucaa.in}

\author[0000-0003-3096-3062]{Lor\'ant O.\ Sjouwerman}
\altaffiliation{Adjunct Professor at the Department of Physics and Astronomy, University of New Mexico.}
\affiliation{National Radio Astronomy Observatory, Pete V. Domenici Science Operations Center, Socorro, NM 87801, USA}
\email{lsjouwer@nrao.edu}

\author[0000-0001-6147-3360]{Anupam Bhardwaj}
\affiliation{Inter-University Centre for Astronomy and Astrophysics (IUCAA), Post Bag 4, Ganeshkhind, Pune 411007, India}
\email{anupam.bhardwaj@iucaa.in}


\begin{abstract}
The structure and evolution of the Milky Way (MW) can be traced with distance estimates to the evolved stellar populations in the inner galactic region. However, direct astrometric distances remain unavailable or highly uncertain for the majority of these sources due to instrumental limitations, large angular diameters, and complex variability. In this work, we develop a supervised machine-learning model to estimate statistical distances to oxygen-rich asymptotic giant branch (AGB) stars selected from the \textit{AKARI} mid-infrared survey. We build an XGBoost regression model that maps multi-band IR photometry to distance using a training set of AGB stars with previously derived SED-based distances achieving a mean absolute percentage error (MAPE) of $\sim6\%$ on an independent test set. Distance estimates for over 36,000 AGB sources are obtained within a 36\% total error margin, greatly expanding distance coverage (0.5--20 kpc) for dust-obscured AGB populations in the Galactic plane. We find good agreement with reported distances for Galactic Mira variables and independent period-luminosity relations. Utilizing the expanded distance set, we further investigate the spatial distribution of Mira variables in the Galactic bulge and disk. Longer-period Miras preferentially trace the bulge's barred morphology compared to their shorter-period counterparts that populate the disk. We also find roughly constant, but differing, relative scale heights for the bulge and disk, with the bulge vertical dispersion about 30\% larger. These findings show that IR photometry-derived statistical distances can recover large-scale Galactic structures and establish long-period Mira variables as efficient tracers of stellar populations in heavily obscured regions of the MW.
\end{abstract}


\section{Introduction}
Understanding the formation and evolution of the Milky Way (MW) requires accurate characterization of the stellar populations that inhabit its inner regions \citep{schultheis2017inner}. Asymptotic giant branch (AGB) stars represent a short but influential evolutionary phase of low- and intermediate-mass stars and are abundant in the inner Galaxy. Through intense mass loss ($10^{-7}$--$10^{-4}\,M_{\odot}\,\mathrm{yr}^{-1}$), AGB stars contribute substantially to the chemical enrichment of the Galactic bulge and disk, returning nuclear-processed material and dust grains to the interstellar medium \citep{Hofner2018}. Despite their high intrinsic bolometric luminosity, identifying these stars at optical wavelengths is challenging due to severe interstellar extinction toward the Galactic Plane, Bulge and Center (GC), compounded by visual obscuration from their own circumstellar envelopes (CSE).

Mid-infrared (mid-IR) surveys have therefore played a critical role in revealing evolved stellar populations in the inner MW. The Mid-course Space eXperiment (MSX)
\citep{Egan_MSX1999} enabled a significant improvement in the identification of variable AGB stars, particularly Mira variables \citep{Sjouwerman_2009}, relative to earlier Infrared Astronomical Satellite (IRAS) color-color selections. Building on this foundation, the Bulge Asymmetries and Dynamical Evolution (BAaDE) assembled a large and homogeneous sample of AGB stars in the inner Galaxy. The BAaDE catalog contains 28{,}062 IR-color selected MSX sources, the majority of which are long-period variables in the inner MW, and provides accurate line-of-sight velocities for well over 10{,}000 objects \citep{Stroh_2019,lewis2021probing}. These maser-bearing Oxygen-rich (O-rich) AGB stars are particularly effective tracers of the Galactic bulge and bar, as their IR brightness and characteristic colors allow them to be identified even in regions of extreme visual extinction \citep{lewis2021probing}.

Reliable distance estimates are essential for such samples to provide useful constraints on Galactic structure and evolution. While optical astrometric surveys such as \textit{Gaia} have revolutionized distance measurements for large stellar populations \citep{Vallenari_gaia_2023}, parallaxes for AGB stars in particular remain difficult to measure and limit the reach to nearby populations. Their faintness in the optical, their large angular diameters, variability-induced astrometric jitter, and complex circumstellar environments all contribute to large parallax uncertainties or systematic biases even for nearby ($d<2$kpc) AGB stars \citep{chiavassa2020optical, lewis2021probing,andriantsaralaza2022distance}. As a result, direct astrometric distances are unavailable or unreliable for the majority of AGB stars in the inner MW.

The limitations of direct astrometry motivate the use of complementary distance estimation techniques, albeit with large distance uncertainties. Kinematic distances, inferred from line-of-sight velocities and Galactic rotation models, suffer from large uncertainties in regions of non-circular motion and complex inner-Galaxy dynamics that is, even after the velocity is successfully and accurately measured. Period-luminosity (P-L) relations provide an alternative, but established calibrations often become unreliable for the dustiest AGB sub-types, where mass-loss rate, metallicity and variability distort observed magnitudes and pulsation properties \citep{Whitelock_pl_2008,Lewis2023}. These challenges highlight the need for distance estimators that can exploit photometric (i.e., IR) data while remaining robust to extinction and variability, for example, by utilizing the spectral energy distribution (SED) \citep{Bhattacharya2024}. We note that these independent complementary methods deriving distance estimates to AGB stars generally suffer from large systematic and random measurement errors, with maximum of about 30-35\%. Adding a different independent method yielding a similar large error therefore may seem fruitless; however any additional reasonable distance estimate has its value when applied in combination with estimates from other methods, or as a proxy in case all other alternatives fail.

Given the lack of reliable distances for distant, highly obscured, and variable AGB stars (described in Sect.~\ref{sec:data}), we develop a supervised machine-learning (ML) regression model that maps multi-band IR photometry to the distance. The model is trained on a large sample of BAaDE/MSX AGB stars with independently derived, physically motivated SED–based distances \citep{Bhattacharya2024}. By learning the empirical relationship between raw, apparent IR photometry and distance, this approach enables statistical distance estimation for sources lacking direct astrometric measurements, without requiring explicit accurate extinction corrections or time consuming flux monitoring programs for the full parent catalog. The here presented ML method provides distance estimates between 0.5- 20 kpc.

The methodology adopted to construct and validate the ML regression model is described in Sect.~\ref{sec:methodology}. Model performance and the resulting distance catalog are presented in Sect.~\ref{sec:results}. In Sect.~\ref{sec:discussion}, we compare the derived distances with Gaia parallaxes and independent P-L relations. We also investigate the spatial distributions of Mira variables in the bulge and disk. The main conclusions of this work are summarized in Sect.~\ref{sec:summary_conclusions}.

\section{Data}\label{sec:data}

We identify a large, clean sample of O-rich AGB candidates across the Galactic plane by applying a color criteria in the mid-IR using \textit{AKARI} 9$\mu$m to 18$\mu$m ([9]-[18]) bands and cross-matching with near-IR (\textit{2MASS}) counterparts. A small fraction of Carbon-rich (C-rich) stars and young stellar objects (YSOs) may remain as impurities, but their statistical weight on the overall population trends is negligible for this study. The data selection and cleaning process are described in detail below.

\subsection{Motivation for Transition from MSX to AKARI}

The BAaDE survey previously identified a sample of $\sim$25\,000 O-rich AGB stars. That sample was constructed primarily using a color-cut in the MSX A--D (8$\mu$m to 15$\mu$m) color, which would maximize the selection of red giants likely to host SiO masers \citep{Sjouwerman2017, Lewis2020}.

Although the MSX color selection effectively identified SiO-emitting AGB stars, it suffered from two key limitations.  
First, the MSX photometric sensitivity declines sharply beyond $\sim$4.5~magnitudes at 8~$\mu$m, the MSX A-band. Apart from $\sim$15 narrow longitude ranges surveyed at higher sensitivity, this limits selection to the brightest and most nearby AGB stars and excludes much fainter, more distant sources (i.e., larger than a few kpc).  
Second, the coarse 18\arcsec\ beam introduces severe source confusion in the crowded inner Galaxy, particularly within $|l|<5^{\circ}$, reducing completeness toward the GC and the far side of the bulge.

To address these issues, we transitioned to the all-sky IR survey conducted by the \textit{AKARI} satellite\footnote{For matching the MSX color selection at a higher angular resolution we determined AKARI to be the most suited. We may expand this study to complimentary surveys like GLIMPSE, WISE, ISOGAL, etc., in the future.} \citep{murakami2007infrared,ishihara2010akari}, designed to provide deeper and higher-resolution coverage in the mid and far-IR compared to the IRAS and MSX surveys.  
\textit{AKARI} comprises two instruments: the \textit{Far-IR Surveyor} (FIS) and the \textit{IR Camera} (IRC).  
  
The IRC, which is most relevant for this work, observed 844{,}649 sources in a 9~$\mu$m band and 194{,}551 sources in an 18~$\mu$m band, with many sources detected in both\footnote{Source counts were taken from the AKARI/IRC Point Source Catalog column description page at IRSA: \url{https://irsa.ipac.caltech.edu/data/AKARI/gator_docs/akari_irc_colDescriptions.html}. This table is available in Vizier too.}
These IRC bands correspond closely to the MSX A (8.28~$\mu$m) and D (14.65~$\mu$m) bands but reach fluxes $\sim$2-3~mag fainter, with higher signal-to-noise ratios and finer angular resolution (5\arcsec-6\arcsec\ FWHM; see Appendix~A of \citealt{ishihara2010akari}). We would also like to note that although the AKARI/IRC all-sky survey products use larger virtual pixels of $\sim$9-10\arcsec, the two-row survey-mode reconstruction preserves a spatial resolution closer to the 5-6\arcsec\ PSF FWHM. Both the virtual pixel scale and reconstructed PSF are substantially finer than the $\sim$18\arcsec\ angular resolution of MSX.  
This improvement substantially reduces source confusion and allows detection of dusty, mass-losing AGB stars several kpc farther into the Galactic bar and bulge and closer toward the direction of the GC ($|l|\ll5^\circ$).

\subsection{AKARI Color Criteria and Source Selection}\label{sec:selection}

We acquired photometric data with high photometric quality ($FQUAL=3$) in both AKARI-9 and -18 bands. Furthermore, to identify a larger, fainter population of O-rich AGB candidates, we employ a conservative color criterion using AKARI bands [9]-[18] by selecting sources lying within $-0.6 < [9]-[18] < 0.6$. We note that using the same AKARI zero-magnitude flux densities as \citet{ishihara2011galactic}, our flux-ratio selection of $-0.6$ to $+0.6$ corresponds to $1.0<[9]-[18]<2.3$ magnitude, consistent with the O-rich AGB region shown in their color-color diagrams.

This color range effectively isolates O-rich, mass-losing AGB stars while minimizing contamination from YSOs and planetary nebulae (Weller et al., in preparation). To further restrict the sample to sources associated with the inner Galaxy, we additionally required Galactic latitudes of $|b| < 5^\circ$, i.e., to match the MSX footprint. We also assessed the contamination level by cross-matching our sample with SIMBAD. After removing duplicate matches and excluding sources with ambiguous classifications, we identified 29,029 sources, corresponding to nearly \(\sim80\%\) of the final sample, with unambiguous SIMBAD classification. Based on these classifications, non-AGB contamination is only \(\sim2.4\%\), dominated primarily by YSO candidates whereas C-rich AGB stars represent a small fraction of the sample, \(\sim0.29\%\). Hence our sample has a very high purity of \(\sim97.6\%\) pure.

\begin{figure}[t]
    \centering
    \includegraphics[width=\columnwidth]{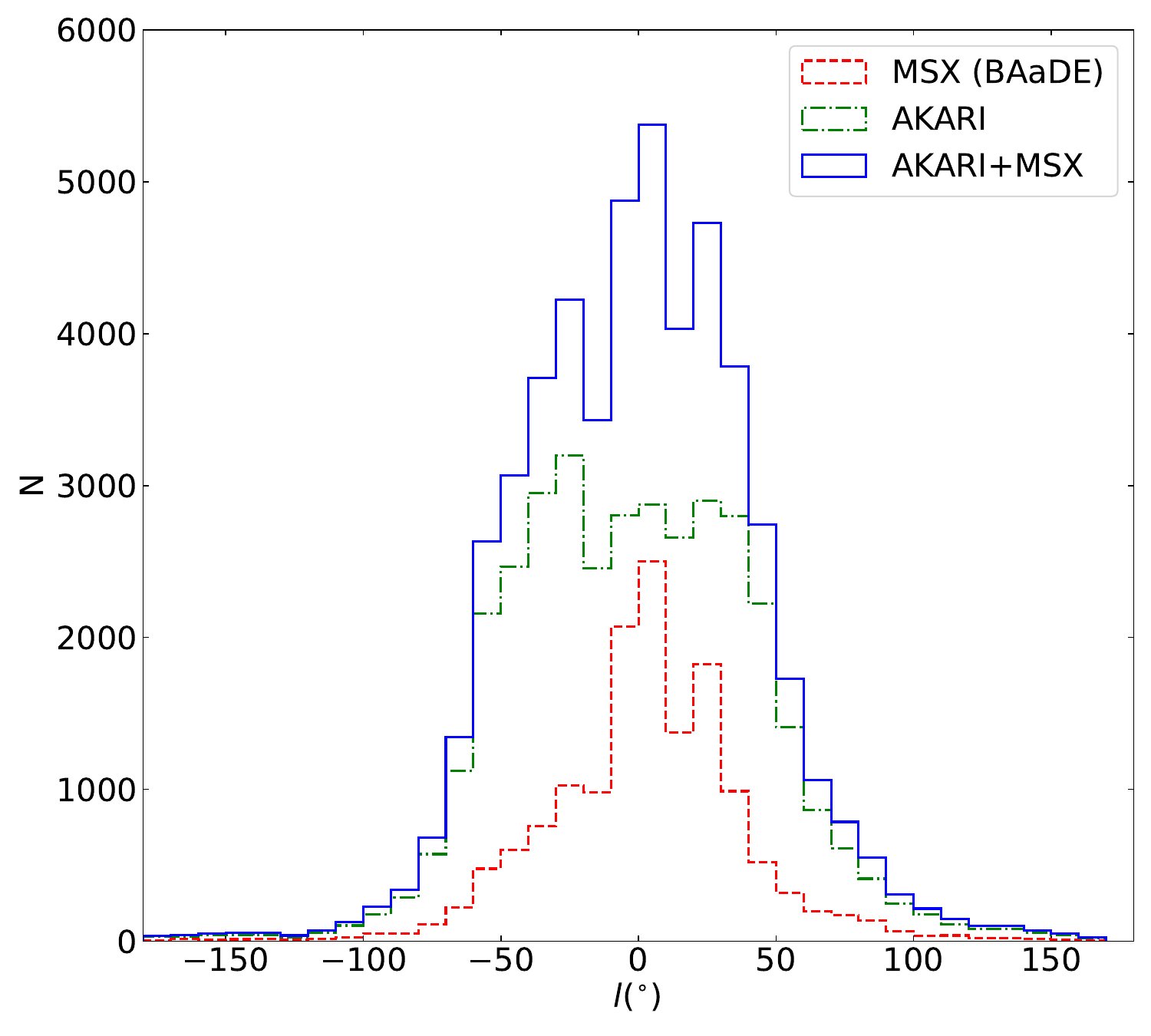}
    \caption{Histogram showing the distribution for galactic longitude ($l$) in degrees of the SED-distance (MSX) sample, the AKARI sample and the combined sample used in this study. Note that the AKARI sample is restricted within $|b|\leq 5^\circ$ to match the MSX/BAaDE footprint.} 
\label{fig:akaripos}
\end{figure}

Applying these boundaries to the full \textit{AKARI}/IRC Point Source Catalog yielded 52{,}767 sources with colors similar to the MSX A-D color criteria used in BAaDE, 
nearly doubling the sources originally in the MSX selection. Figure~\ref{fig:akaripos} shows the Galactic longitude distribution of the MSX (BAaDE) sources for which distances were obtained in the earlier analysis of \citet{Bhattacharya2024}; hereafter, we refer to these sources as the \textit{MSX-selected sample}. From the distribution, we see that the \textit{AKARI} sample provides substantially more coverage of both the near and far sides of the Galactic bulge, whereas the MSX sample shows a pronounced deficit of far-side sources ($-30\lesssim l^\circ \lesssim 0$). Consequently, the \textit{AKARI}-selected sample displays a broader and more symmetric longitude distribution about the GC, consistent with the improved sensitivity and resolution of \textit{AKARI}, enabling detections of fainter AGB stars and supposedly extending the accessible volume several kpc farther into the bulge. Furthermore, the improved depth and angular resolution of \textit{AKARI}, in particular in combination with the MSX sample, provides a significantly larger sample of AGB stars suitable for distance and population analyses. 

\begin{figure*}[t]
    \centering
    \includegraphics[width=\textwidth]{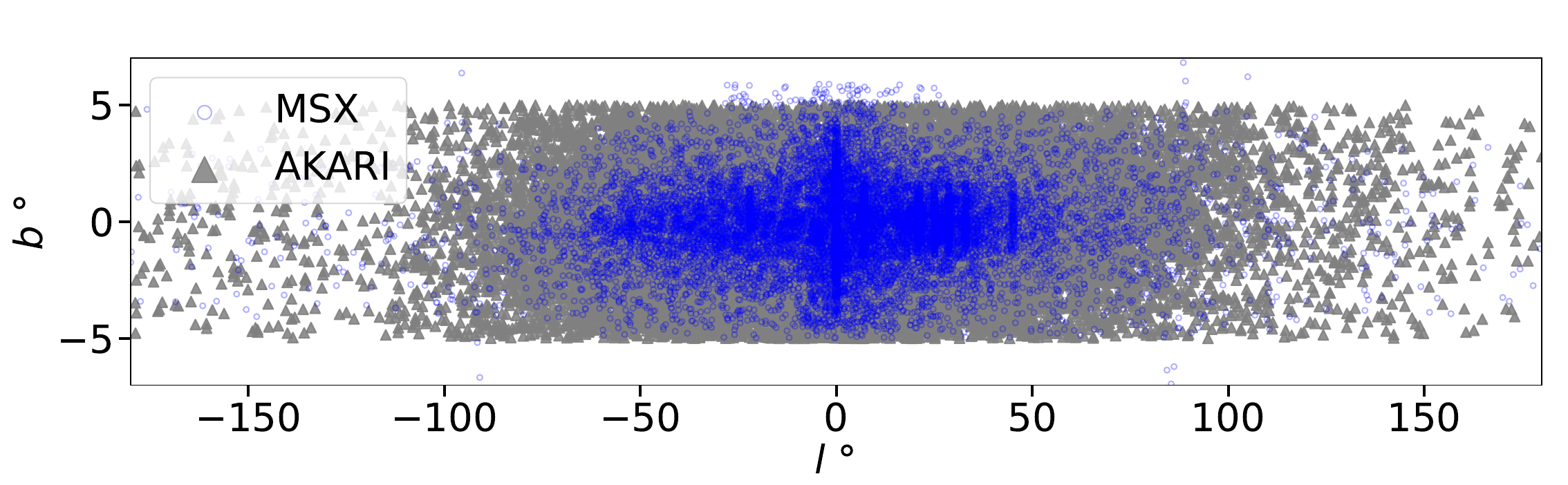}
    \caption{Sky distribution of the MSX-selected and AKARI-selected samples in Galactic coordinates $(l,b)$. The MSX-selected sample is shown as blue open circles, while the AKARI-selected sample is shown as gray triangles. The AKARI sample extends over a broader longitude range than the MSX sample, while both samples occupy a similar low-latitude Galactic footprint. This comparison complements the longitude histogram in Fig.~\ref{fig:akaripos} and illustrates the improved sky coverage provided by the AKARI selection.}
\label{fig:lbdistribution}
\end{figure*}

To complement the longitude histogram in Fig.~\ref{fig:akaripos}, we also show the sky distribution of the MSX-selected, AKARI-selected, and combined samples in Galactic coordinates in Fig.~\ref{fig:lbdistribution}. This figure confirms the same overall trend seen in the longitude histogram: the AKARI-selected sample provides broader longitude coverage than the MSX-selected sample, particularly in the far side. Both samples by design occupy a similar Galactic latitude range, although the more distant AKARI-selected sources may in principle experience larger cumulative interstellar extinction because their sightlines traverse longer path lengths through the Galactic disk.

\subsection{Cross-matching with 2MASS} \label{cm}

To associate the mid-IR \textit{AKARI} detections with near-IR photometry, we cross-matched the \textit{AKARI} selection with the \textit{2MASS} Point Source Catalog \citep{cutri2006explanatory} using the CDS X-Match service. For this, we chose a search radius of 1\arcsec\ as AKARI has a better angular resolution ($\sim$ 5\arcsec) compared to the  angular resolution of MSX (18\arcsec) and better matches the $\sim$ 1\arcsec\ angular resolution of \textit{2MASS}. This also ensured removal of the redundant sources in the sample, and for the multiple matches, we considered the reddest sources. To ensure high photometric reliability, only sources with a quality flag of ``AAA'' in the \textit{2MASS} catalog were retained. This procedure ensured uniform and high-quality $JHK_s$ photometry across the dataset (see Table \ref{tab:data_summary}) and eliminated spurious or blended detections. 
Further, from the total AGB candidate sample (47,216 sources), we exclude the overlapping MSX-selected (BAaDE) sample (11,082 sources), as the latter is utilized for training and validation of the ML model (for details, see Sect.~\ref{sec:methodology}). As a result, the final AGB candidate sample consists of 36{,}134 sources (hereafter referred to as the ``\textit{AKARI} AGB sample") on which the ML model is employed to estimate the distances and assess spatial trends across the Galactic bar (see Sect.~\ref{subsec:model_application}).

\begin{deluxetable*}{lcc}
\tablecaption{Summary of Source Counts After Each Selection Step}
\label{tab:data_summary}
\tablehead{
\colhead{Selection Step} & \colhead{Criterion Applied} & \colhead{Number of Sources}
}
\startdata
\textit{AKARI} [9]-[18] color selection & $-0.6 < [9]-[18] < 0.6$, $FQUAL=3$,  $ |b|\lesssim 5^\circ$ & 52,767 \\
Cross-match with \textit{2MASS} & 1\arcsec\ radius, Quality flag ``AAA'' & 47,216 \\
Cleaned sample (removed SED overlaps) & independent of MSX (BAaDE) SED-distance catalog & 36,134\\ 
\enddata
\tablecomments{The \textit{AKARI} [9]-[18] color criteria yield nearly twice as many color-selected candidates as the MSX A--D selection. 
Subsequent crossmatching with \textit{2MASS} provides near-IR photometry.}
\end{deluxetable*}

\section{Methods}\label{sec:methodology}

In this study, we estimate statistical distances for the \textit{AKARI} AGB sample by training a supervised ML model on the available MSX-selected sample from \citet{Bhattacharya2024}. By combining these previously published distances with an eXtreme Gradient Boosting (XGBoost) regression framework \citep{Chen_2016}, we establish an empirical mapping between multi-band IR photometry and stellar distance. This approach enables us to extend SED-based distance estimates to the full \textit{AKARI} AGB sample. In the following subsections, we examine the data sample and briefly describe ML model development and implementation.

\begin{figure}[t]
    \centering
    \includegraphics[trim=0.0cm 0.0cm 0.0cm 0.0cm,clip, width=\columnwidth]{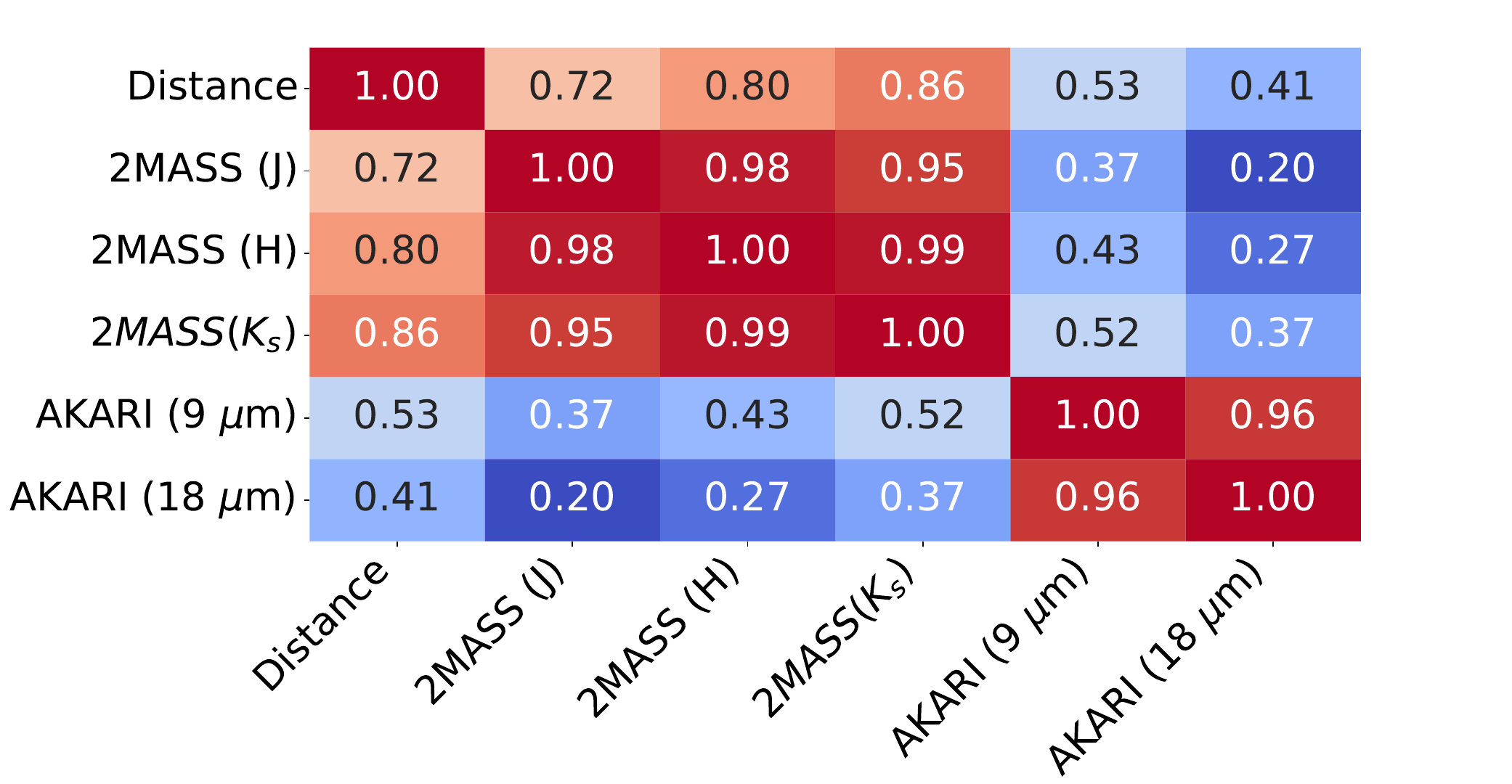}
    \caption{Correlation matrix between the non–extinction-corrected IR photometric 2MASS and AKARI bands and the SED–derived distances for the MSX-selected sample (training $+$ test set).}
    \label{fig:correlation_matrix}
\end{figure}

\begin{deluxetable*}{cc}
\tabletypesize{\scriptsize}
\tablewidth{0pt} 
\tablecaption{Best values for key parameters of the XGB distance regressor.\label{tab:ML_model_params}}
\tablehead{
\colhead{Model Parameters} & \colhead{Values}
} 
\startdata 
n\_estimator & 360 \\
max\_depth & 7 \\
learning\_rate & 0.03144 \\
min\_child\_weight & 4 \\ 
sub\_sample & 0.66957 \\
\enddata

\end{deluxetable*}

\subsection{Feature definition and preprocessing}\label{subsubsec:features}

We use infrared photometry to construct a feature vector for each source in the training (and testing) set (i.e the MSX-selected sample with available SED distance), which comprises of 14,654 objects.\footnote{Of these, 11,082 sources overlap with the AKARI AGB sample and were removed from our catalog, as described in Section~\ref{cm}. However, these sources are retained in our training set.} 

Further, 4,542 sources from the MSX-selected sample with missing data in any of the five 2MASS and AKARI IR bands are dropped, resulting in 10,112 sources. We then randomly divide the final training set into an 80\% training and a 20\% test set. The training set is used for model fitting and hyperparameter optimization, while the test set is used only for model evaluation. Additionally, for the hold-out test, we use a $K$-fold cross-validation procedure on the training set to test model robustness and ensure that performance is not sensitive to the train–validation partition. The feature vector includes apparent magnitudes from 2MASS ($J$, $H$, $K_{s}$) and AKARI (9\,$\mu$m, and 18\,$\mu$m) bands, which is available for all of the SED-distance and main AKARI samples. We find strong correlations between the apparent IR photometry (without any de-reddening) and the SED-derived distances (Fig.\,\ref{fig:correlation_matrix}), especially for the near-IR bands, which exhibit Pearson correlation coefficients of $r \simeq 0.7-0.9$ with distance, indicative of strong predictive strength of the magnitudes towards distance estimation. 

Although the AKARI 9 and 18~$\mu$m bands exhibit weaker individual correlations with distance than the near-infrared bands, they provide complementary information that enhances the overall predictive performance when used alongside near-infrared features. In tree-based ensemble methods, performance depends on the availability of informative predictors, and excluding relevant features can limit the model’s ability to capture underlying structure, leading to underfitting \citep{hastie2009elements}. The AKARI bands, which trace circumstellar dust emission, therefore provide complementary information that improves the characterization of the spectral energy distribution and enhances the accuracy of the ML-based distance estimates. We also adopt apparent magnitudes as model inputs, as applying extinction corrections to the full \textit{AKARI} AGB sample ($\sim 36{,}000$ sources) would require extensive ancillary data and could introduce additional systematic uncertainties.

We would like to note that the ML-derived distances should be interpreted as empirical distances tied to the SED-based distance scale of the training set. They are not expected to be more accurate than the SED-based distances themselves, because the model performance is ultimately limited by the uncertainties and systematics of the input training distances. The advantage of the ML approach is instead its ability to provide homogeneous statistical distance estimates for a much larger sample of AGB stars, enabling improved statistical mapping of the inner Galaxy.

A direct SED-matching approach similar to \citet{Bhattacharya2024} would require suitable calibration templates with independently constrained distances, such as sources with VLBI parallaxes or sources inferred to lie near the Galactic Center from their line-of-sight velocities and Galactic coordinates. This information was available for the MSX sample, but line-of-sight velocities are not available for most AKARI-selected sources. The ML approach helps bypass these limitations by learning the empirical relation between multi-band infrared photometry and the SED-based distance scale using the AKARI photometry of the MSX training set. In this sense, the training set transfers the distance calibration from the 2MASS-MSX SED templates to the 2MASS-AKARI photometric space. The AKARI-selected sources occupy a near- and mid-IR color-magnitude space similar to that of the MSX-selected training sample, so applying the model to the AKARI catalog does not require a large extrapolation beyond the parameter range represented in the training set. The reliability of this application is further supported in Sects.~\ref{sec:results} and \ref{sec:discussion}, where we compare our predicted distances with other available distance estimates and find statistically consistent behavior, with no evidence for a systematic bias specific to our predicted distances.

\begin{deluxetable*}{lcc}
\tabletypesize{\scriptsize}
\tablewidth{0pt}
\tablecaption{Performance metrics of the XGBoost distance--regression model. \label{tab:ml_performance}}
\tablehead{
\colhead{Dataset} &
\colhead{$R^{2}$} &
\colhead{MAPE (\%)}
}
\startdata
Training set          & 0.981             & \nodata \\
Test set              & 0.980             & 94.06  \\
5-fold CV (mean)      & $0.963 \pm 0.003$  & \nodata \\
\enddata
\tablecomments{
The 5-fold cross-validation row reports the mean and standard deviation across the folds.
}
\end{deluxetable*}

\subsection{XGBoost regression model}\label{subsubsec:xgb_model}

Using the \texttt{XGBRegressor} implementation in the \texttt{scikit-learn} ecosystem, we model the nonlinear relationship between IR photometry and stellar distance. In XGBoost, the prediction is the sum of many sequentially trained gradient-boosted decision trees. The model iteratively refines its target function approximation by fitting each new tree to the ensemble residuals.

The key model hyperparameters listed in Table \ref{tab:ML_model_params} are: \textit{n\_estimator} (number of trees in the ensemble), \textit{max\_depth} (maximum tree depth, managing model complexity), \textit{learning\_rate} (shrinkage parameter regulating tree influence),   \textit{min\_child\_weight} (splitting of tree nodes when instance weight falls below a given value), and \textit{sub\_sample} (stochastically determine training sub-sample). Regularization and leaf weights reduce model complexity and overfitting. The model is optimized using a differentiable loss function. XGBoost quickly converges using first- and second-order loss function derivatives.

\subsubsection{Model training and optimization}\label{subsubsec:training}

Model development is carried out using the 80\% training subset of the MSX sample. Within this training set of about 8\,000 sources, we adopt a $K$-fold cross-validation scheme (with $K=5$) to assess model stability and mitigate overfitting. At each iteration, the data are split into $K$ folds; the model is trained on $K-1$ folds and validated on the remaining fold, and this process is repeated so that each fold serves as the validation set once. The cross-validated performance metric is the root mean square error (RMSE) between the predicted and input SED-derived distances.

Tuning of the model hyperparameters is performed using \texttt{optuna} framework \citep{akiba2019optuna}, a highly efficient Bayesian optimization strategy. During each trial, the optimizer selects a new hyperparameter set, trains the \texttt{XGBRegressor} using 5-fold cross-validation, and records the mean RMSE. After full trials, the optimal model configuration is the hyperparameter combination with the lowest cross-validated RMSE. The complete set of tuned hyperparameters is listed in Table~\ref{tab:ML_model_params}.

Using the best hyperparameters, we retrain the \texttt{XGBRegressor} on the entire training set. Training is stopped on an internal validation subset from the training data when the validation loss does not improve after a certain number of boosting rounds. This prevents the model from growing beyond optimal generalization.

\begin{figure}[t]
    \centering
    \includegraphics[width=\columnwidth]{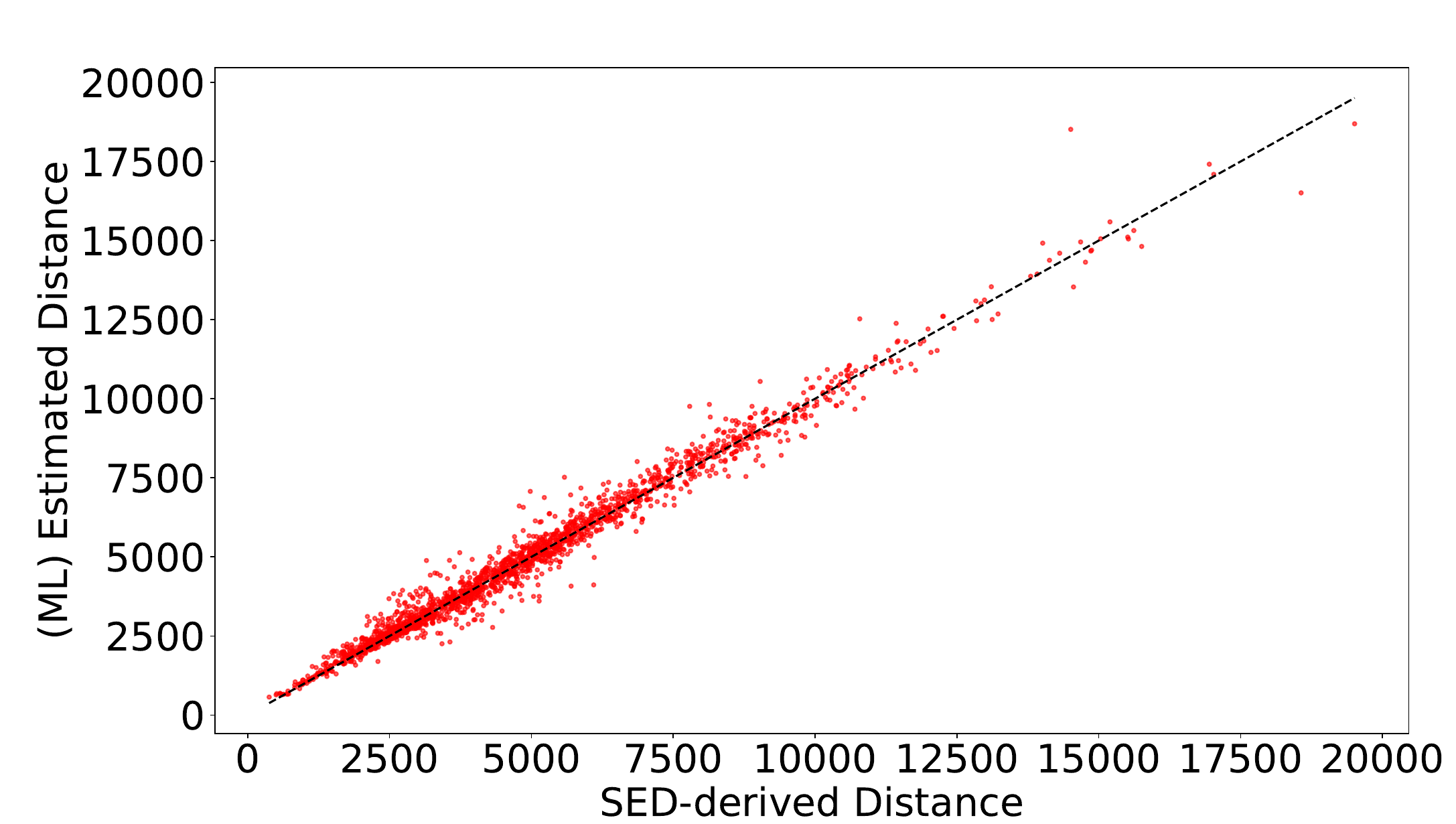}
    \caption{ML-estimated or predicted distances for the test set versus the input original SED-derived distance \citep{Bhattacharya2024}. Red points show individual sources and the dashed line marks the one-to-one relation. The close alignment indicates strong model performance with minimal scatter across the full distance range.}
    \label{fig:finaldist}
\end{figure}

\subsubsection{Model evaluation}\label{subsubsec:evaluation}

The final model is evaluated using the remaining 20\% test set, unused during training or hyperparameter optimization. We quantify performance using standard regression metrics, such as the coefficient of determination ($R^{2}$), the mean absolute error (MAE), and the RMSE. We also report a complementary accuracy metric based on the mean absolute percentage error (MAPE), expressed as
\begin{equation}
    \mathrm{Accuracy}_{\rm MAPE} = 100 - \left( \frac{100}{N} \sum_{i=1}^{N} 
    \left| \frac{\hat{y}_i - y_i}{y_i} \right| \right) \%,
\end{equation}
where $y_i$ and $\hat{y}_i$ denote the original SED-derived and ML-predicted distances, respectively, and $N$ is the number of test sources ($\sim 2100$). This gives a sense of the typical fractional deviation between model predictions and SED distances. 

\section{Results}\label{sec:results}
In this section, we outline the performance of the model on both the training and test sets and evaluate the model's stability during the training phase. Following this, we apply the model to the whole \textit{AKARI} AGB sample to predict the distances and trace their distribution in the MW.

\subsection{ML model performance}\label{subsec:global_performance}
The model's ability to accurately predict the distances is judged from both training and test sets. We analyze the model's robustness and stability using a variety of evaluation parameters, including regressor metrics (explained in Sect.~\ref{subsubsec:evaluation}). Furthermore, we assess the model's performance throughout the boosting process across the entire range of distances associated with AGB sources.

\subsubsection{Final performance metrics}
The full set of quantitative metrics is summarized in Table~\ref{tab:ml_performance}. It is noted that the training and test $R^{2}$ values are essentially the same ($R^{2} \approx 0.98$), indicating that the model captures the dominant variance in the mapping between IR photometry and SED-derived distances without noteworthy overfitting. The MAE and RMSE likewise show close agreement between the training and test samples, confirming that the model generalizes effectively to previously unseen test data. Therefore, the optimized \texttt{XGBRegressor} model demonstrates strong and consistent performance across both the training and test sets.

To provide a complementary, interpretable measure of fractional accuracy, we compute a MAPE-based accuracy metric for the test sample. The resulting value of $\sim94\%$ corresponds to the mean deviation in the distance estimates by $\sim6\%$. The predicted versus original SED-based distances for the test set (Fig.~\ref{fig:finaldist}) follow a tight one-to-one correlation with minimal systematic deviations. We note the same behavior across the full distance range probed by the BAaDE sample (0.5-20 kpc).

Additionally, during training, we evaluate the stability of the model from the individual fold results across the training set. The individual fold results (Table~\ref{tab:ml_performance}) show a narrow dispersion in $R^{2}$, with a mean of $0.963$ and a standard deviation of only $0.003$. The corresponding MAE remains similarly stable, showing a variation of less than~1\% across the folds. These results confirm that the model performance is not sensitive to any specific train–validation split and that the established multi-band photometry–distance relation is statistically well constrained. A close agreement between the $R^{2}$ regression metric for cross-validation and test-set performance suggests an unbiased regression model.

\subsubsection{Diagnostic behavior of the regression model}\label{subsec:diagnostics}
\begin{figure}[t]
    \centering
    \includegraphics[width=\columnwidth]{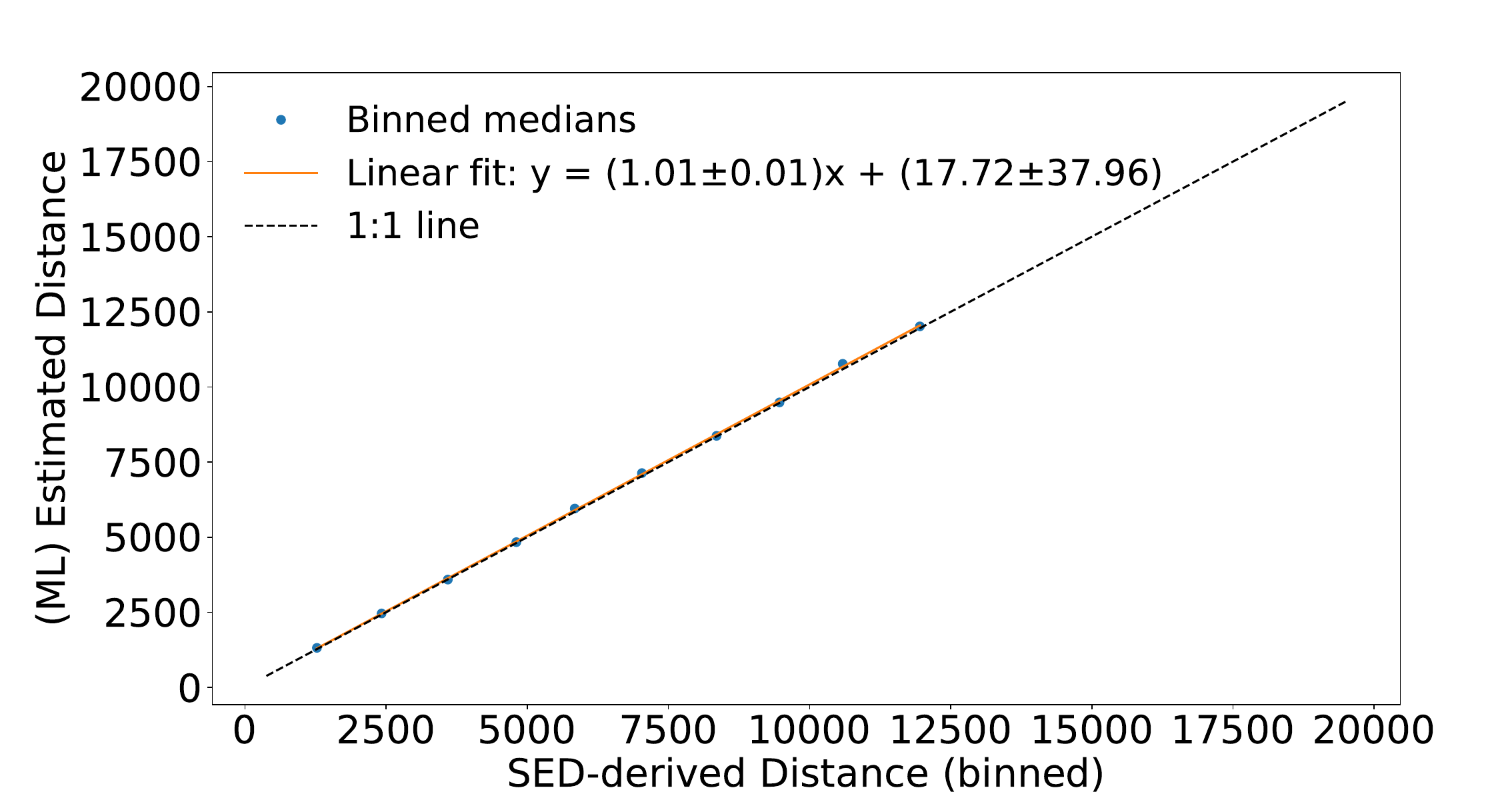}
    \caption{Comparison of ML-predicted distances with SED-based distances for the test sample. Individual sources are binned in two dimensions to reduce scatter in high-density regions. The black dashed diagonal line represents the 1:1 relation, while the blue points show the binned median trend, indicating the 16th-84th percentile spread within each bin. The close agreement between the median trend and the 1:1 line, along with the relatively small scatter, demonstrates that the model accurately reproduces the SED-based distances across the full range of the sample.}
    \label{fig:predmedian}
\end{figure}

\begin{figure}[t]
    \centering
    \includegraphics[width=\columnwidth]{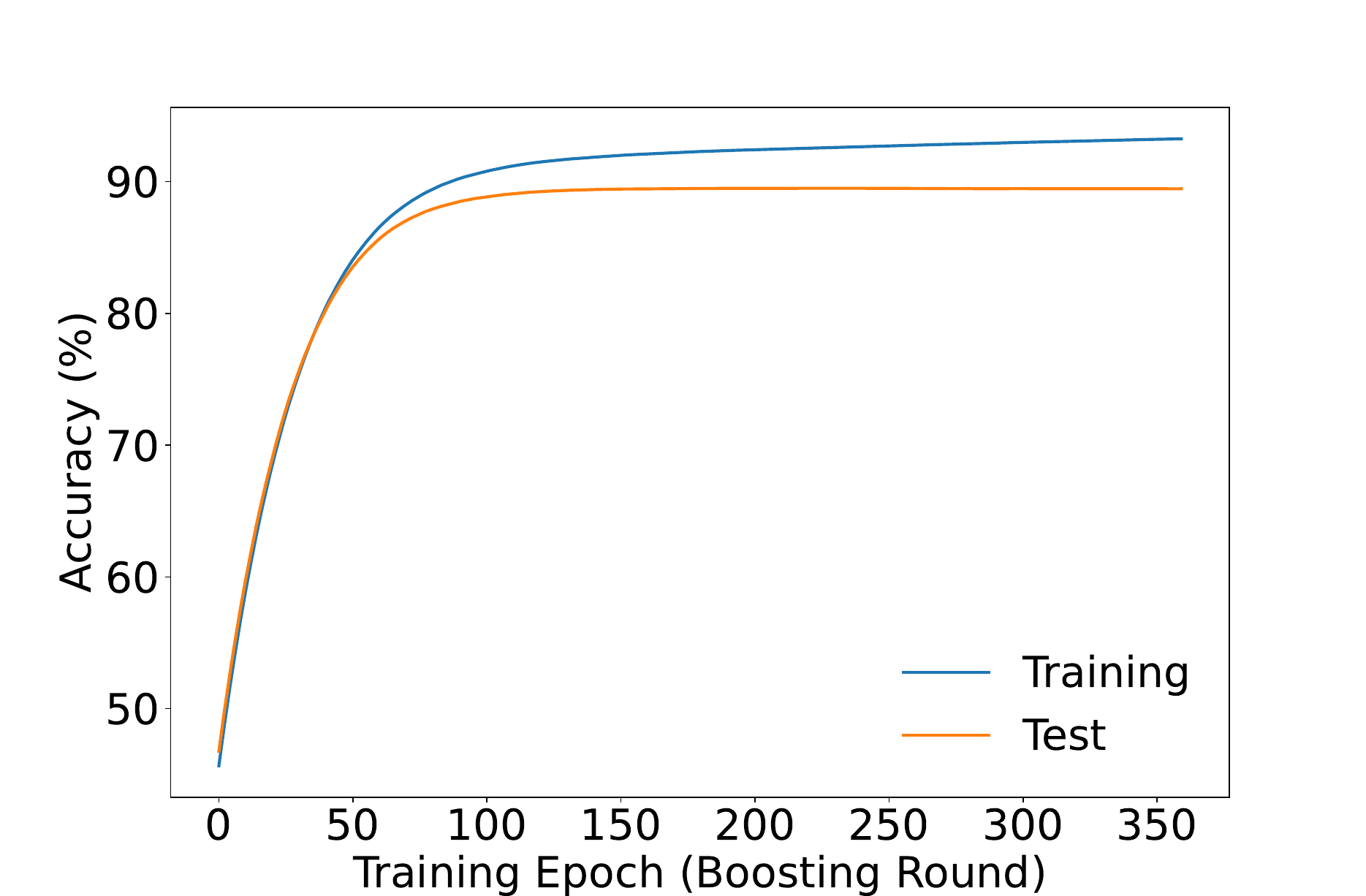}
    \caption{Training and test accuracy as a function of boosting rounds for the \texttt{XGBRegressor} model. The curves display how model performance evolves over the course of training, with early boosting rounds producing rapid improvements followed by convergence toward a stable plateau. The close agreement between the training and test accuracy curves indicates minimal overfitting and demonstrates that the model generalizes well to unseen data. This diagnostic behavior confirms that the chosen hyperparameters yield a well-regularized model suitable for distance prediction.}
    \label{fig:boosting}
\end{figure}

A crucial step involves assessing the \texttt{XGBRegressor} model for potential instability or overfitting during the prediction of stellar distances. To evaluate the fidelity and stability of the model, we examine two diagnostic schemes as stated below:\\
(a) The relation between the model-predicted and SED-based distance is shown in Fig.~\ref{fig:predmedian}. We compare the median binned ML-predicted distances to the SED-derived distance, with both quantities binned, and overplot the median relation (including the 16th-84th percentile bands). The relation between the model-predicted and SED-based distance is shown in Fig.~\ref{fig:predmedian}. 
\begin{equation}
d_{\mathrm{ML}} = (1.01 \pm 0.01)\, d_{\mathrm{SED}} + (17.72 \pm 37.96)\,\mathrm{pc}
\end{equation}
which is consistent with a near-unity slope and negligible offset. With no discernible scatter or curvature, the predicted distances exhibit a close 1:1 relationship with the input SED distances over the entire range where there are a sufficient number ($N \geq 10$) of sources per bin. This behavior demonstrates that the model performs robustly from nearby objects to sources several kpc away. The narrow percentile envelope further indicates that the scatter is well controlled. 
(b) The evolution of training and test accuracy as a function of boosting round (see Fig.~\ref{fig:boosting}) is tracked. It is observed that both the training and test accuracies follow a smooth curve and approach their peak value just above 100 boosting rounds. In addition, the gap between them remains small during boosting, indicating that the model avoids overfitting and that the hyperparameters produce a stable and well-regularized solution. The accuracy curve plateau confirms that the model reaches an optimal condition before test performance may degrade well beyond 350 boosting rounds of the optimal model.

Therefore, results from both these diagnostics clearly show that the model remains accurate and unbiased across the full parameter space of the MSX sample.

\subsubsection{Summary of model performance}\label{subsec:summary_results}

The results from the performance statistics and diagnostics demonstrate that the \texttt{XGBRegressor} model provides model-accurate, stable, and largely unbiased distance estimates using a handful of IR photometric data. With $R^{2} \approx 0.98$ for both training and test sets and typical fractional errors of $ \sim6\%$, the ML model distances offer a reliable extension of the SED-based distance framework. This enables distance estimation for the full \textit{AKARI} AGB catalog, supporting large-scale analyses of Galactic structure, kinematics, and evolved stellar populations in the inner dust-obscured regions of the MW as shown in Sect.\, \ref{subsec:model_application} which is further discussed in Sect.\,\ref{sec:discussion}.

\subsection{Model application on the AKARI AGB sample}\label{subsec:model_application}

After training and validating the \texttt{XGBRegressor} model on the MSX-selected sample, the model is applied on the complete catalog of 36,134 \textit{AKARI} AGB candidates to estimate their distances using the same set of near- and mid-IR photometric data listing results in Table \ref{tab:pred_agb}. In the following subsections, we examine the spatial distribution of the AGB sample along with the magnitude of error in their predicted distances.

\begin{deluxetable}{cccr}
\tablecaption{ML-predicted (statistical) distances including the estimated 1$\sigma$ errors for the \textit{AKARI} AGB sample (see Sect.~\ref{subsec:model_application}). Only the first 10 rows are shown. The full table is available electronically.
\label{tab:pred_agb}}
\tablehead{
\colhead{Object} & \colhead{RA (J2000)} & \colhead{Dec (J2000)} & \colhead{Distance}\\
(AKARI ID) & (hh:mm:ss) & (dd:mm:ss) & (kpc)
}
\startdata
0026205+621627 & 00:26:20.534 & 62:16:27.437 & $8.6 \pm 3.1$ \\
0034301+664135 & 00:34:29.960 & 66:41:35.106 & $5.4 \pm 1.9$ \\
0045196+583018 & 00:45:19.696 & 58:30:18.389 & $18.8 \pm 6.7$ \\
0105302+582626 & 01:05:30.263 & 58:26:26.290 & $4.6 \pm 1.6$ \\ 
0125475+614200 & 01:25:47.586 & 61:42:00.990 & $3.8 \pm 1.3$ \\
0137539+644448 & 01:37:54.157 & 64:44:48.354 & $9.2 \pm 3.3$ \\
0213070+625719 & 02:13:07.194 & 62:57:19.037 & $8.9 \pm 3.2$ \\
0219342+582135 & 02:19:34.242 & 58:21:35.053 & $11.4 \pm 4.0$ \\
0302323+620307 & 03:02:32.306 & 62:03:07.934 & $5.0 \pm 1.8$ \\ 
0315124+601131 & 03:15:12.628 & 60:11:30.228 & $4.2 \pm 1.5$ \\
\enddata

\end{deluxetable}

\subsubsection{Galactic distribution of the AGB candidates}

\begin{figure}[t]
    \centering
    \includegraphics[width= 0.5\textwidth]{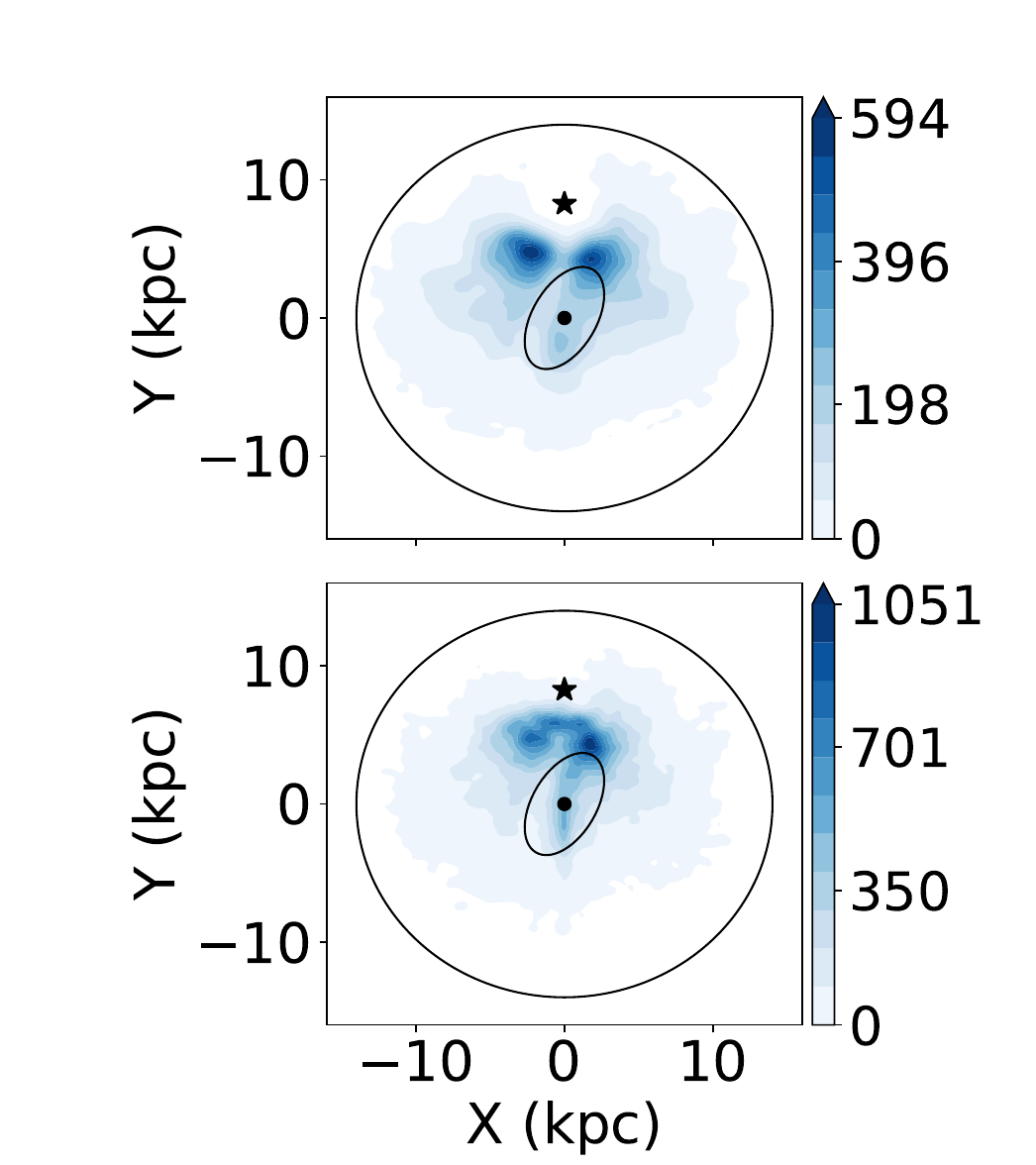}
    \caption{Projected two-dimensional $(X,Y)$ density maps of AGB candidates in the Galactic plane (i.e., $|b|<5^\circ$). The Galactic Centre (GC) is marked by a filled circle at $(0,0)$, and the Sun is indicated by a star at $(0, 8.277~\mathrm{kpc})$. An ellipse centered on the GC, with a semi-major axis of 4~kpc and a semi-minor axis of 2.2~kpc, inclined by $27^\circ$ clockwise relative to the Sun-GC direction, outlines the approximate extent of the Galactic bulge following \citet{wegg2015structure}. The colorbar indicates the effective number of sources contributing locally. \textit{Top:} Density map of the \textit{AKARI-only} AGB sample alone. \textit{Bottom:} Density map of all sources from \textit{AKARI} AGB and MSX sample. The colorbar indicates source density in the $(X,Y)$ plane, in units of $N~\mathrm{kpc}^{-2}$. where N is  number of sources.}
    \label{fig:densitydist}
\end{figure}

We investigate the spatial distribution of all the \textit{AKARI} AGB candidates and how they trace the inner MW’s large-scale structure using the projected two-dimensional $(X,Y)$ density maps shown in Fig.~\ref{fig:densitydist}. The \textit{top panel} displays the distribution of the \textit{AKARI-only} AGB candidates while the \textit{bottom panel} shows the combined distribution of the \textit{AKARI} AGB candidates together with the MSX-selected sample. We note that the sensitivity of the surveys emphasize the sources in the foreground, but looking beyond that, in the \textit{AKARI}-only map, an elongated overdensity is evident in the inner Galaxy, with an orientation broadly consistent with expectations for a barred Galactic bulge. When the MSX-selected sample is included (bottom panel), the overall bar-like morphology becomes even more prominent. The apparent connection between the near-side density enhancements in the combined map is largely a consequence of the different longitude coverage, and treatment of overlapping sources in the two samples. As shown in Fig.~\ref{fig:akaripos}, the AKARI-selected sample provides broader longitude coverage, including more sources toward the far side of the bulge, whereas the MSX-selected sample is more concentrated toward Galactic-center longitudes in particular including a more sensitive (deep) field at $l=0^\circ$. In addition, a subset of sources common to the MSX-selected and AKARI samples was removed from the final AKARI catalog to avoid duplication, but these sources are retained in the MSX-selected training and comparison sample (See Sect.\ \ref{subsubsec:features}). Therefore, when the MSX-selected sample is added to the AKARI sample, additional sources at relatively small heliocentric distances fill the region between the near-side density enhancements, making the projected density distribution appear more continuous. A similar projected structure is also visible in the $(X)$-$(Y)$ distribution of AKARI O-rich AGB stars presented by \citet{ishihara2011galactic}, suggesting that this morphology is already present in the AKARI-selected AGB catalog and is not introduced by our distance-estimation method. This supports the interpretation that the apparent clumps are primarily induced by the AKARI selections and its  longitude coverage. Therefore, the enhanced densities are most likely caused by longitude-dependent source sampling rather than by a distinct physical Galactic component.

To assess whether the Galactic bar signature is present, we further investigate the distance distributions of all sources (\textit{AKARI}+MSX), restricting the sample to two inner-Galaxy longitude intervals, $8^\circ < \ell < 16^\circ$ and $-12^\circ < \ell < -4^\circ$ and also within $|b| < 2^\circ$ to remove foreground sources. These intervals are intentionally asymmetric in longitude in order to probe comparable Galactocentric radii along the bar on either side of the Galactic center. Although including all sources along the line of sight introduces substantial contamination from foreground disk populations and broadens the resulting distributions, a clear systematic offset remains: the distribution of objects at positive longitudes peaks at smaller heliocentric distances than that at negative longitudes.

\begin{figure}[t]
    \centering
    \includegraphics[width= \columnwidth]{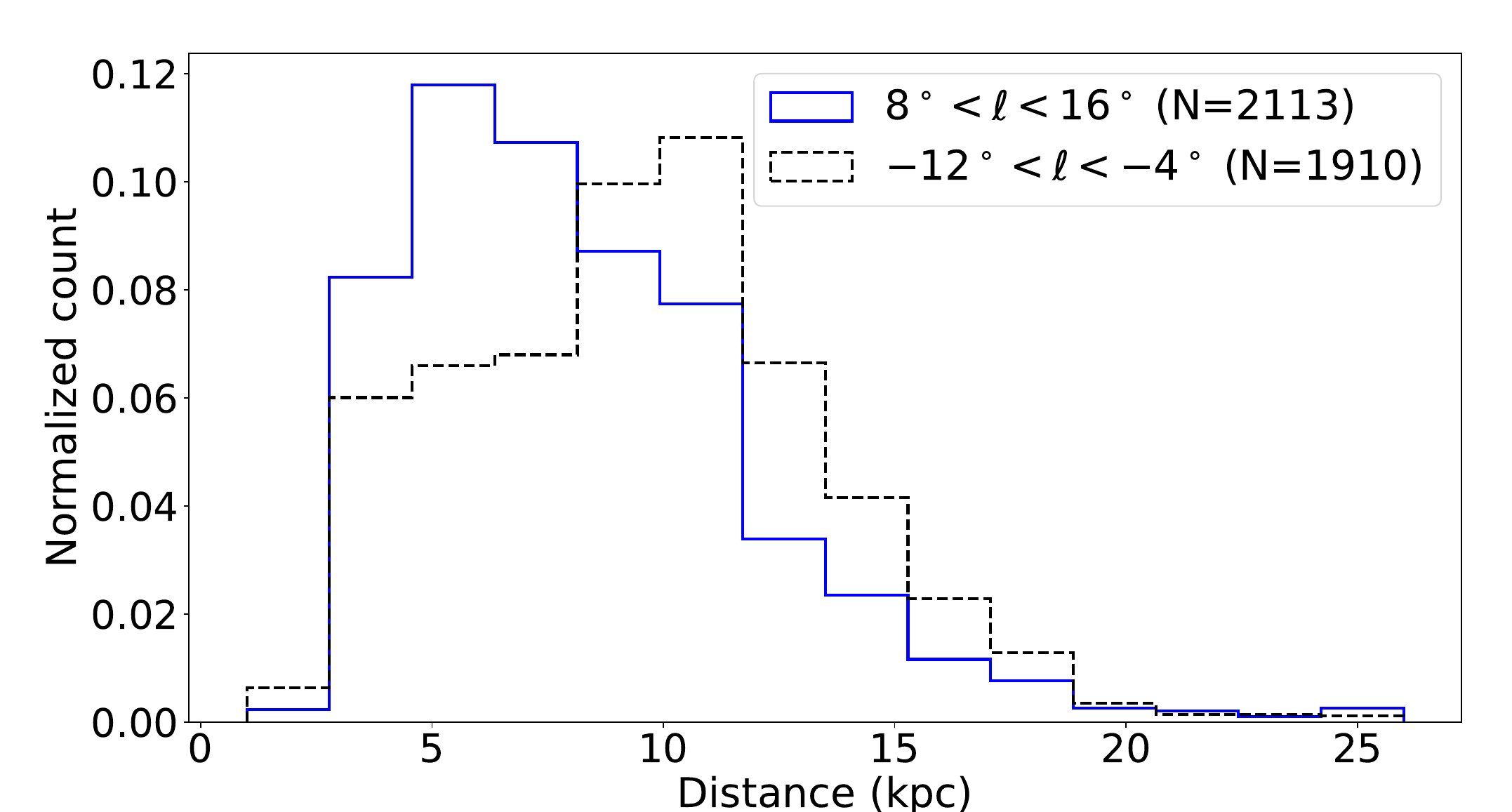}
    \caption{Histogram highlighting the distance-based population distribution of the combined \textit{AKARI} AGB sample and MSX-selected sources for the near ($8^{\circ}<l<16^{\circ}$) and far-side ($-12^{\circ}<l<-4^{\circ}$) of the MW. }
    \label{fig:bulgehist}
\end{figure}

The resulting distance distributions (Fig.~\ref{fig:bulgehist}) reveal a distinct near-far separation, with near-side peaks occurring at smaller distances and far-side peaks at larger distances. If the bulge were spherically symmetric, the peak of the distance distributions toward positive and negative longitudes would be statistically indistinguishable. The observed offset therefore implies a non-axisymmetric structure and is consistent with the geometric signature of an inclined, triaxial bar whose near end lies at positive Galactic longitudes. This interpretation is consistent with previous Mira-based studies of the Galactic bulge. Previous studies using AGB populations have also reported a triaxial, bar-like structure in the Galactic bulge \citep[e.g.,][]{whitelock1991iras,whitelock1992shape,catchpole2016age}. In the context of these earlier results, our analysis provides an independent consistency check, showing that the ML-derived AKARI distances recover the expected bar-like structure traced by evolved AGB/Mira populations. We further discuss this comparison and its implications in more detail in Sect.~\ref{subsec:variab}.

Despite the lack of explicit dynamical modeling in this work, the agreement between the observed distance asymmetry and the expectations for barred geometry indicates that the bar signature remains detectable even without imposing an explicit bulge selection, supporting the use of the \textit{AKARI} AGB sample as an effective tracer of the inner Galaxy.

\subsubsection{Error budget of our distance estimates}
It is noted that the MSX (BAaDE) SED-distances that are used for model development have a maximum uncertainty of $\sim35\%$, which could arise from factors such as circumstellar extinction and photometric variability \citep{Bhattacharya2024}. Further in Sect.\ \ref{subsubsec:evaluation}, we note that the ML model introduces an additional methodological uncertainty with an average value of $\sim6\%$. Combining both these uncertainties in quadrature yields a representative total maximum uncertainty $\sim35.5\%$ which is adopted for all the sources in Table \ref{tab:pred_agb}. Although we validate our predictions against several independent distance estimates (in Sect.~\ref{sec:discussion}), we emphasize that the resulting distances should still be regarded as statistical rather than accurate measurements for individual objects. While individual stars may exhibit larger uncertainties, the ensemble of distances remain well constrained and suitable for population-level analyses. As a result, regardless of the range of uncertainty in the individual ML-predicted distances, the estimates can be utilized to clearly trace the bulge as well as both the far and near sides of the MW.

\section{Discussion}\label{sec:discussion}

\subsection{Comparison with other distance estimates}\label{subsec:comparison_distance}

To evaluate the reliability of the ML-predicted distances, we compare them with independent distance estimates derived from \textit{Gaia} parallaxes and established P-L relations similar to \citet{Bhattacharya2024}. We would like to note that each of these methods has its own limitations and potentially large errors on individual sources, and therefore provides complementary constraints on the validity of our approach. 

\subsubsection{Comparison with Gaia parallaxes} \label{subsec:gaiaplx}

Independent validation of distance estimates for heavily obscured AGB stars remains challenging. \textit{Gaia} distances for AGB stars are known to be systematically underestimated beyond $\sim2~\mathrm{kpc}$ \citep{2025A&A...698A.109L}. Even when restricting to high-quality parallaxes with reported fractional uncertainties below 20\%, converting a parallax to a distance introduces significant additional uncertainties for these stars \citep{andriantsaralaza2022distance}. Consequently, apart from a limited number of nearby and well-measured cases, \textit{Gaia}-based distances do not provide a robust benchmark for the present sample. 

\begin{figure}[t]
    \centering
    \includegraphics[trim=0.0cm 0.0cm 0.0cm 0.0cm,clip,width=\columnwidth]{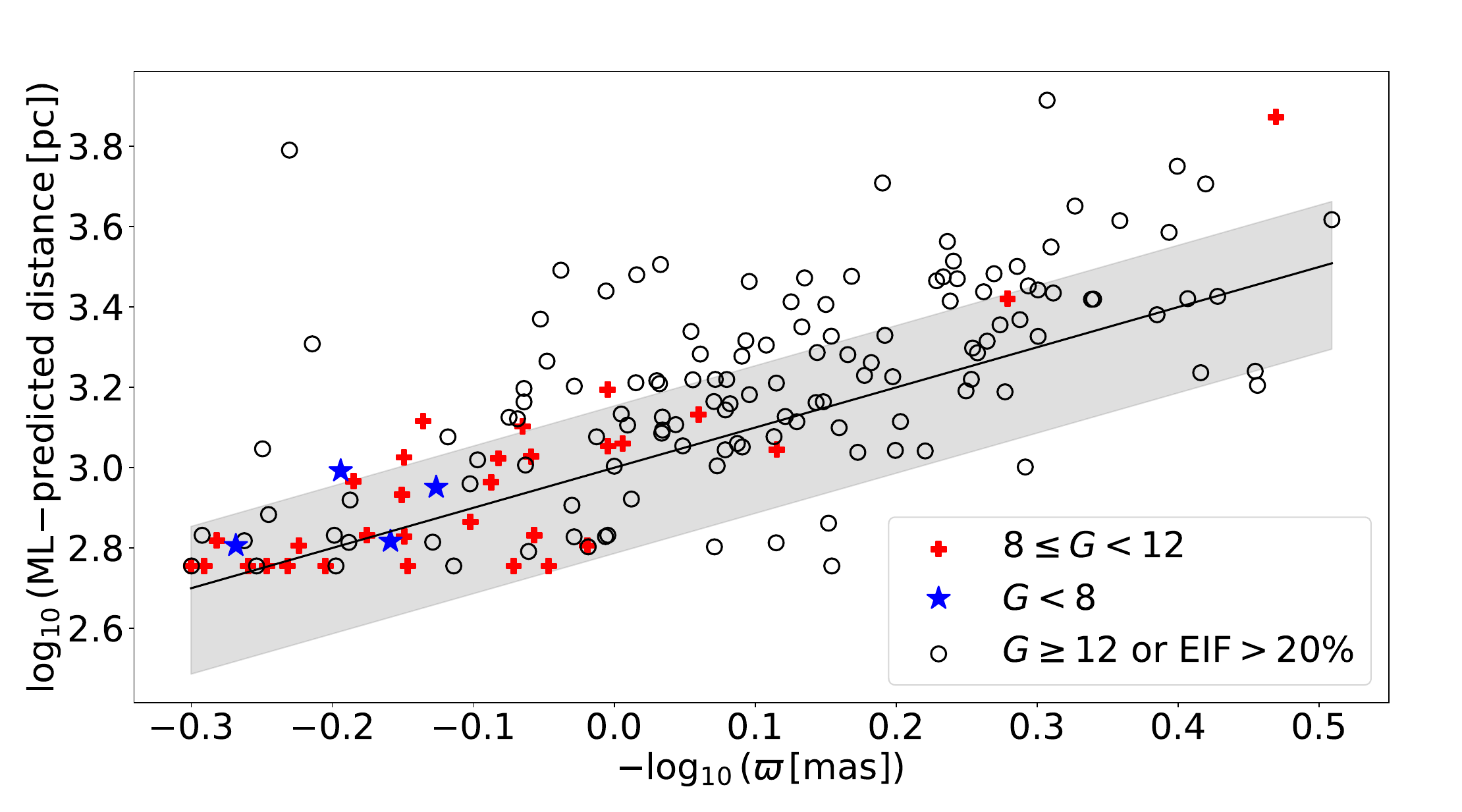}
    \caption{Comparison between ML-predicted distances and \textit{Gaia} parallaxes for 166 cross-matched sources with $|b|> 5^\circ$. The ML-predicted distances (in pc) are shown in logarithmic scale and plotted against $-\log_{10}(\varpi)$, where $\varpi$ is the \textit{Gaia} parallax in mas. In this representation, larger values along the horizontal axis correspond to smaller parallaxes (i.e., larger distances). The blue symbols indicate the brightest \textit{Gaia} sources with $G<8$, while the red symbols correspond to sources with $8<G<12$ for which the EIF-corrected fractional parallax uncertainties are below $20\%$. Sources with $G>12$ or with EIF-corrected fractional parallax uncertainties exceeding $20\%$ are shown as open circles. The solid line represents 1-1 correlation. The shaded region indicates the combined uncertainty envelope, corresponding to $20\%$ fractional uncertainty in parallax and $35.5\%$ fractional uncertainty in the ML-predicted distances, propagated in logarithmic space. We restrict the sample to sources with predicted distances greater than $0.5~\mathrm{kpc}$, consistent with the range over which the model is applicable and note that the \textit{Gaia} distances are systematically closer compared to the ML-derived distances, in particular for the fainter sources ($G>12$).}
  \label{fig:martidist}
\end{figure}

Although we attempted to cross-match the \textit{AKARI} AGB sample with \textit{Gaia}, no sources within the \textit{AKARI} sample (restricted to $|b| < 5^\circ$, also see Sect.~\ref{subsec:lumgaia}) have reliable \textit{Gaia} parallaxes (they are all in the MSX sample that was removed from the \textit{AKARI} sample). As an external consistency check, we instead utilize sources from \citet{marti2025gaia} at higher Galactic latitudes ($|b| > 5^\circ$, outside the MSX footprint), for which \textit{Gaia} parallaxes are available. For these objects, that all have good \textit{AKARI} and 2MASS photometry, we apply our ML model to estimate distances and compare them with the corresponding parallax-based distances. Out of that sample, 166 sources have associated \textit{Gaia} DR3 parallaxes with fractional uncertainties $<20\%$, for which parallax-based distances can be estimated in a straightforward manner \citep{bailer2015estimating}. However, AGB stars are known to exhibit large uncertainties in \textit{Gaia} astrometry due to their strong variability, extended atmospheres, and circumstellar obscuration \citep{Xu_2019}. \citet{andriantsaralaza2022distance} further demonstrate that even sources with nominal parallax uncertainties below $20\%$ can have significantly underestimated errors, and introduce an EIF that depends on the (apparent) $G$ magnitude, with the largest corrections required for the brightest sources ($G<8$~mag).

Figure \ref{fig:martidist} shows the relation between ML-predicted distances and \textit{Gaia} parallaxes in logarithmic space. The ML-predicted distances broadly follow the expected inverse-parallax trend. The overall Spearman rank correlation coefficient of $\rho = -0.72$ confirms a strong monotonic inverse correlation (distance correlated with the inverse of parallax). The slope of $-1$ between the ML-predicted distances and \textit{Gaia} parallaxes indicates this consistency with inverse relationship. The brightest sources ($G<8$) show the tightest correspondence, while sources with intermediate magnitudes ($8<G<12$) exhibit increased dispersion. The faintest sources ($G>12$), for which no EIF correction is available, display the largest scatter.

A systematic offset between the ML-predicted distances and those implied by parallaxes becomes more apparent toward lower parallaxes (larger distances), with the ML estimates tending to exceed the corresponding parallax-based distances. This behavior is consistent with the known tendency of \textit{Gaia} parallaxes to underestimate distances beyond $\sim 2~\mathrm{kpc}$ due to biases such as the Lutz–Kelker effect. The inverse-correlation is stronger for brighter and EIF-corrected sources, indicating improved agreement between the ML-predicted distances and the more reliable \textit{Gaia} parallaxes in these regimes. Overall, the comparison demonstrates that the ML-predicted distances are broadly consistent with \textit{Gaia} measurements, while also highlighting the limitations of parallax-based distances from \textit{Gaia} for distant and heavily obscured AGB stars.

\subsubsection{Stellar Luminosities from Gaia OH/IR catalog}\label{subsec:lumgaia}
We also cross-matched our $|b|<5^\circ$ \textit{AKARI} AGB sample with the Galactic \textit{Gaia} OH/IR star catalog compiled by \citet{marti2025gaia}, identifying 143 counterparts within a matching radius of $5''$. These sources lack reliable \textit{Gaia} parallax measurements. However, the catalog provides estimates of bolometric fluxes ($F_{\rm bol}$), which, when combined with our distance estimates, enable the calculation of bolometric magnitudes for this subsample.

\begin{figure}[t]
    \centering
    \includegraphics[trim=0.0cm 0.0cm 0.0cm 0.0cm,clip,width=\columnwidth]{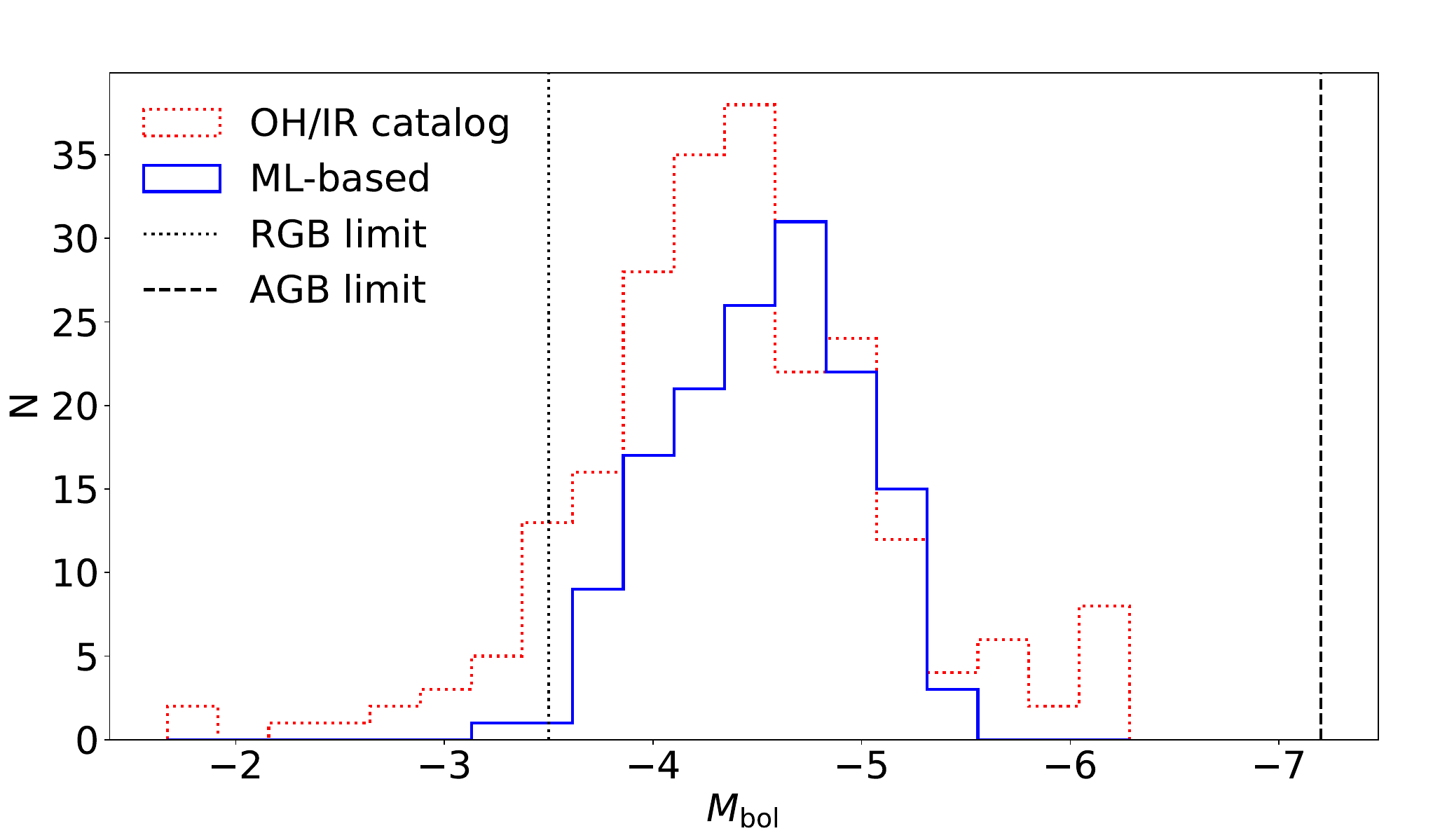}
    \caption{Bolometric magnitude distribution for 143 \textit{Gaia} OH/IR sources in our sample without \textit{Gaia} parallaxes satisfying a fractional parallax error $<20\%$, whose luminosities are derived using ML-based distances, compared with 222 other sources from \citet{marti2025gaia} that possess parallaxes with fractional parallax error $< 20\%$ and have luminosities based on parallax-derived distances following \citet{bailer2015estimating}.}
    \label{fig:martilum}
\end{figure}

Upon comparison with the findings of \citet{marti2025gaia}, we find that the peak of the bolometric magnitude distribution derived using our distance estimates is shifted toward brighter values by $\sim$0.4~mag (Fig. \ref{fig:martilum}). However, the luminosities reported by \citet{marti2025gaia} are primarily based on nearby sources (within $\sim 2$~kpc) with \textit{Gaia} parallaxes with fractional errors $<20\%$ and distances estimated following \citet{bailer2015estimating}.

Although we cross-match with the catalog of \citet{marti2025gaia}, we would like to note that the luminosity distributions are constructed from different subsets. Their luminosity distribution includes sources with reliable \textit{Gaia} parallaxes (and therefore predominantly nearby objects), whereas our luminosity sample consists of sources lacking reliable \textit{Gaia} parallaxes and thus not included in their analysis.

The difference in the peak luminosity may also reflect the underlying stellar mass distribution of the two samples. Since AGB luminosity correlates with initial stellar mass, the inclusion of a larger fraction of more massive stars in our sample—particularly those located in the inner Milky Way—naturally leads to a higher luminosity peak. This suggests that the two samples probe distinct subsets of the Galactic AGB population: the \citet{marti2025gaia} sample predominantly traces nearby, lower-mass (and hence lower-luminosity) AGB stars, whereas our sample includes a greater fraction of more massive, intrinsically luminous sources at larger distances.

As an additional consistency check, we examined the variability properties of the cross-matched sources using the \textit{AAVSO Variable Star Index} (VSX; \citealt{Watson2006}). Of the 143 objects, 
65 are classified as Mira variables with well-defined pulsation periods. Applying the P-L relation of \citet{feast1989period} to these 65 Miras yields a $M_{\rm bol}$ distribution with a slightly higher peak at $\sim -5.0$~mag. The bolometric fluxes used here are not corrected for interstellar extinction, implying that the derived luminosities represent lower limits. Although extinction is lower at $|b| \gtrsim 2^\circ$ compared to the Galactic midplane, it is still significant. Therefore, correcting for extinction should raise the luminosity distribution, improving agreement with \citet{feast1989period} P-L estimates.

Overall, the luminosity distribution derived from our distance estimates is physically plausible and consistent with expectations for AGB stars. Moreover, broad agreement with independent P-L predictions supports the interpretation that the inferred distances yield realistic luminosities at the population level.

\subsubsection{Comparison with P-L estimates}
Given the limitations of \textit{Gaia} parallaxes, we also interpret the inferred distances through comparisons with distances derived from established P-L relations, while recognizing that these methods introduce systematic uncertainties at the level of a few tens of percent for individual sources. These comparisons are limited to sources that are directly cross-matched with samples for which P-L-based distances are available; we do not apply P-L relations to the full \textit{AKARI} sample, as reliable extinction corrections are not available and not all sources have sufficiently accurate periods measured. Nevertheless, the size of the cross-matched samples remains sufficiently large to enable statistically meaningful comparisons.

\begin{figure*}[t]
    \centering
    \includegraphics[trim=0.0cm 0.0cm 0.0cm 0.0cm,clip,width=\textwidth]{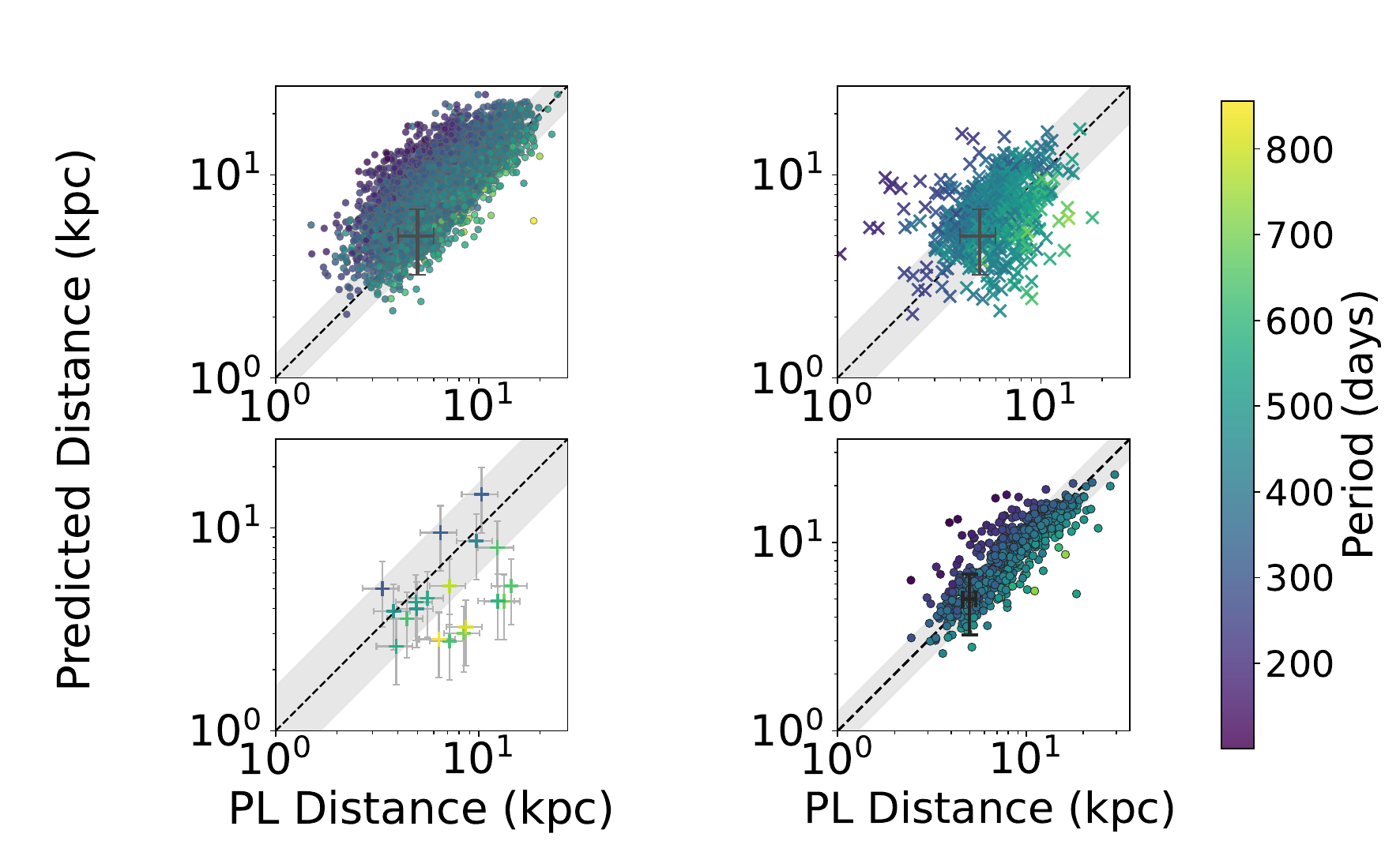}
\caption{
Comparison between ML-predicted distances and P-L-based distances for different cross-matched samples. 
The panels are ordered clockwise from the top-left: OGLE-based near-IR P-L distances from \citet{iwanek2023three}, mid-IR MSX-based P-L distances from \citet{lewis2023long}, $K_s$-band Milky Way Mira P-L distances from \citet{sanders2023period}, and near-IR P-L distances from \citet{urago20203d}. 
The dashed black line marks the one-to-one relation, and the shaded gray regions show the scatter about this relation for each comparison. 
Points are colored by stellar period, and representative error bars are shown for both the P-L and ML-predicted distances.}
\label{fig:akari_pl_sed}
\end{figure*}

Figure~\ref{fig:akari_pl_sed} summarizes these comparisons in a four-panel layout. 
Starting from the top-left panel, we compare the ML-predicted distances with near-IR P-L distances derived from OGLE periods using relations calibrated on Large Magellanic Cloud (LMC) Mira variables \citep{iwanek2023three}, based on 8{,}032 cross-matched sources. A clear systematic offset is observed, increasing with distance in absolute terms, with the near-IR P-L distances being on average 36\% shorter than the ML-predicted distances. This behavior is consistent with known limitations of applying LMC-calibrated near-IR P-L relations to Galactic AGB stars. In particular, the calibration of \citet{iwanek2023three} adopts a Galactic Center distance of 7.66~kpc, about 8\% smaller than the value of 8.277~kpc derived by \citet{abuter2019geometric}, which intrinsically scales distances downward with an error that increases with distance. Additionally, the LMC-based calibration reflects a lower-metallicity environment compared to the predominantly more metal-rich Galactic population probed by \textit{AKARI}.

The top-right panel shows the comparison with mid-IR P-L distances for Galactic Mira variables, derived using the MSX A, C, and D bands and calibrated following \citet{lewis2023long}, based on 608 cross-matched sources. The agreement between the two distance estimates improves relative to the near-IR comparison, with the systematic offset largely reduced ($\sim 12\%$). However, substantial scatter persists at the level of individual sources, reflecting residual uncertainties in both methods.

The bottom-right panel compares the ML-predicted distances with P-L distances based on the calibration of \citet{sanders2023period}, who used refined \textit{Gaia}-parallax distances and sources from the \textit{Gaia} LPV catalog to derive P-L relations for Milky Way Mira variables. Cross-matching their catalog with our sample yields 738 common sources. For these sources, we compute distances using their $K_s$-band O-rich Mira P-L relation. The ML-predicted distances broadly follow the P-L-based distances, with most sources lying close to the one-to-one relation. We also compared our distances with those obtained using the LMC-based relation from \citet{sanders2023period}; in that case, the LMC-based P-L distances are systematically shorter than our ML-predicted distances similar to what we see in the OGLE comparison.

The bottom-left panel presents a comparison with a smaller subsample from \citet{urago20203d}, based on an independent near-IR P-L calibration which was also calibrated using LMC AGB stars. Approximately half of the sources are consistent with the ML-predicted distances within their quoted $1\sigma$ uncertainties. While the limited sample size prevents strong statistical conclusions, this comparison highlights the intrinsic challenges of using P-L relations for distance determination. We further cross-matched our catalog with \citet{albarracin2024unveiling}, but only found 3 common sources. For these few matches, the distances reported by \citet{albarracin2024unveiling} are approximately a factor of two larger than our ML-predicted distances. However, because the number of matches is very small, we do not draw any further conclusions from this comparison.

We note that the ratio of the P-L over ML based distances in all panels in Fig.\ \ref{fig:akari_pl_sed} increase with increasing period (shifting from above to below the 1-1 line). As this is an effect involving the stellar period, we cautiously interpret this with a general issue of determining distances with P-L relations and not due to our ML method, which does not have an explicit dependence on period. The quantitative behavior illustrated in Fig.~\ref{fig:akari_pl_sed} suggests that period–luminosity relations calibrated on Galactic AGB stars, provide a more appropriate reference for interpreting the inferred distances than relations derived from LMC AGB populations. While the distances derived in this work should be regarded as statistical rather than accurate measurements for individual objects, the consistency across independent Galactic calibrations supports their suitability for population-level studies of Galactic structure.

\begin{figure*}[t]
    \centering
    \includegraphics[trim=0.0cm 0.0cm 0.0cm 0.0cm,clip,width=\textwidth]{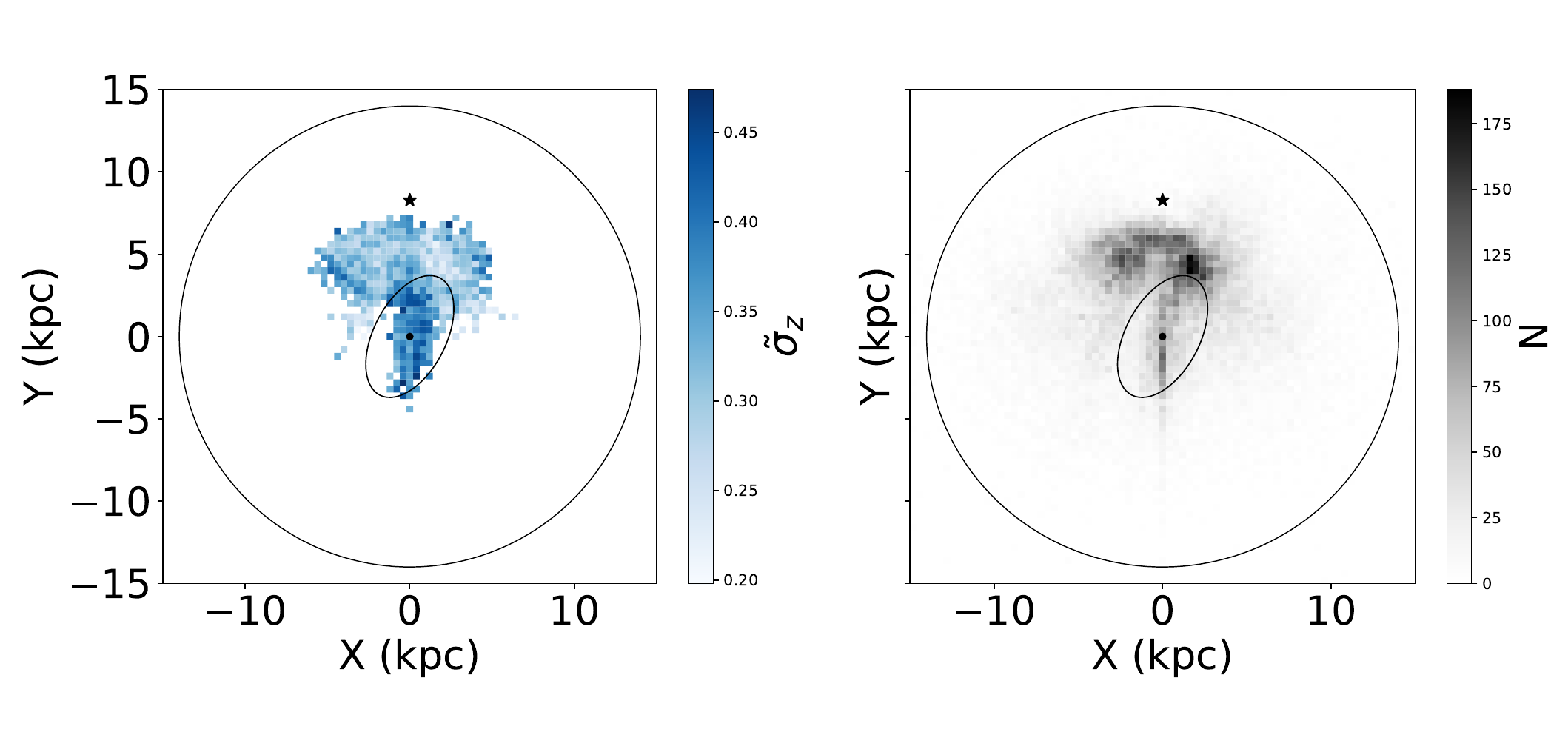}
    \caption{
Heatmap of the dimensionless vertical dispersion, $\tilde{\sigma}_z$, in the Galactic plane. The left panel shows $\tilde{\sigma}_z$ computed in $(X,Y)$ bins, where the colorbar represents the value of $\tilde{\sigma}_z$, while the right panel shows the corresponding number of sources per bin. The Sun is marked at $(0, 8.277)$ and the Galactic center at $(0,0)$. The overlaid ellipse represents the Galactic bar/bulge with semi-major axis $a=4.0$~kpc, semi-minor axis $b=2.2$~kpc, and an orientation angle of $27^\circ$. The map shows enhanced vertical dispersion in the central bulge region compared to the surrounding disk, with a clear transition between the two components. The $(X,Y)$ bin size is 500~pc, and only bins containing at least 35 sources are included in the derivation of $\tilde{\sigma}_z$.
}
    \label{fig:heatmap}
\end{figure*}

\begin{figure}[t]
    \centering
    \includegraphics[trim=0.0cm 0.0cm 0.0cm 0.0cm,clip,width=\columnwidth]{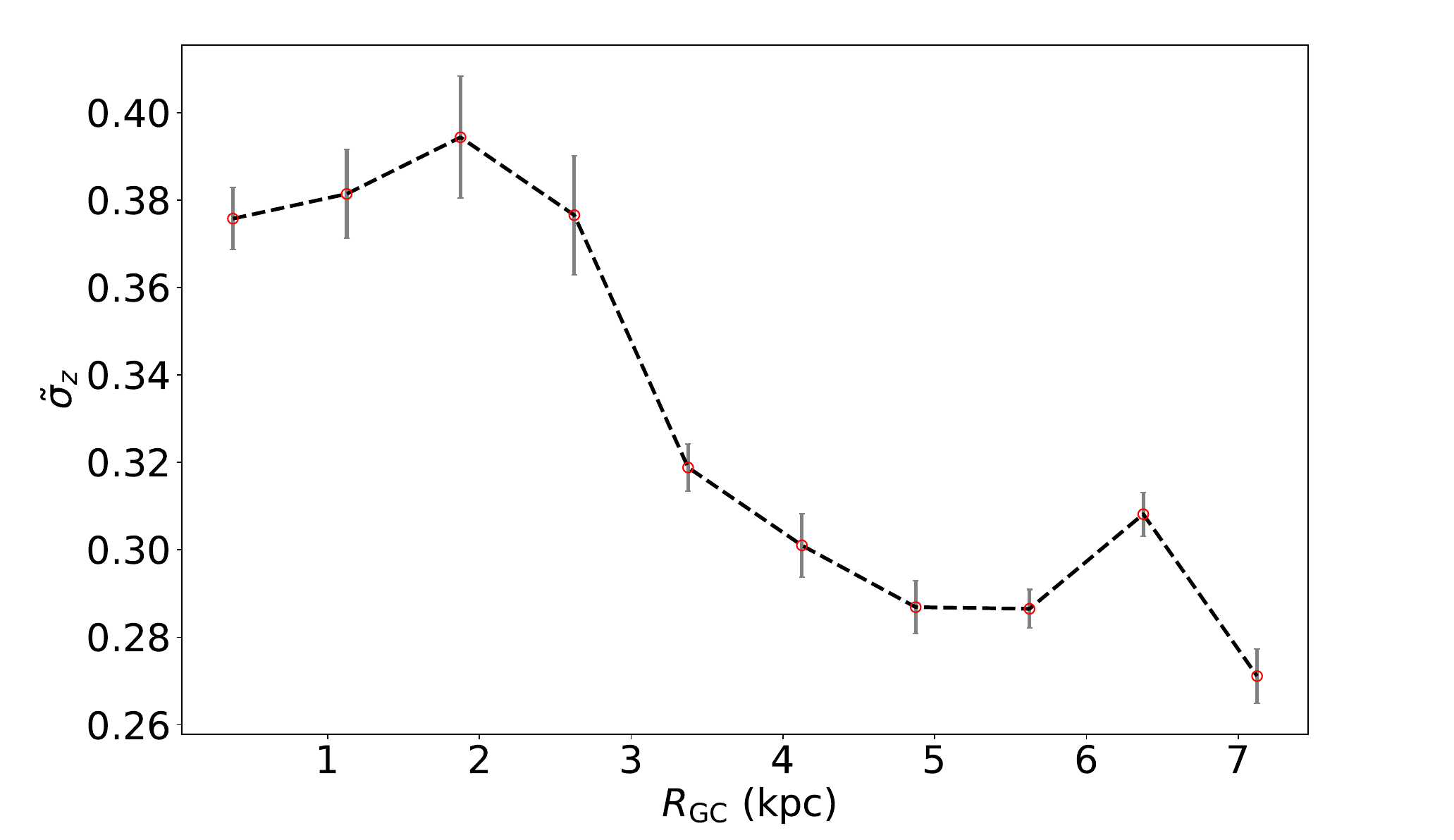}
    \caption{
Radial variation of the dimensionless vertical dispersion, $\tilde{\sigma}_z$, as a function of Galactocentric radius, $R_{\rm GC}$. The values are derived by averaging over spatial bins in the $(X,Y)$ plane. The error bars represent the uncertainty in each radial bin, which depends on the number of sources contributing to that bin. The profile shows a relatively high and nearly uniform dispersion in the inner regions, followed by a decline and flattening toward larger radii, consistent with a transition from the thicker bulge to the  thinner Galactic disk.
}
    \label{fig:dispersionplot}
\end{figure}

We also tested whether interstellar extinction introduces a systematic bias in the ML-predicted distances by comparing the distance residuals with available \(A_{K_s}\) estimates for sources with independent distance measurements from OGLE and mid-IR PLRs. In the central Galactic-plane region, \(-5^\circ < l < 5^\circ\) and \(-2^\circ < b < 2^\circ\), where extinction effects are expected to be strongest, \(A_{K_s}\) does not show a significant correlation with either the absolute or fractional distance residuals. Thus, highly extincted sources are not preferentially assigned larger distances by our method. Therefore, although offsets exist between our distances and some literature distance scales, these offsets are not driven by extinction-dependent systematics in the ML predictions. This also supports the interpretation that the main density structures recovered in this work are not produced by extinction systematically shifting sources to larger inferred distances.

\subsection{Galactocentric scale height dispersion}

We investigate the vertical structure of the inner Galaxy using a combined sample of AGB stars consisting of \textit{AKARI} AGB candidates and the MSX-selected sample. The combined datasets provide a large enough sample and a statistically robust characterization of the vertical dispersion across the inner disk and bulge. The vertical extent of a stellar population is determined by its distribution in height above the Galactic plane, $Z = d \sin b$, where $d$ is the heliocentric distance and $b$ is the Galactic latitude. In the absence of observational constraints, the dispersion in $Z$ would be the same as the intrinsic scale height of the population. However, our sample is subject to a latitude selection, which introduces a geometric truncation in the observable vertical range.

Our sample is limited to Galactic latitudes $|b| \lesssim 6^\circ$, such that the accessible vertical extent at a given distance is bound by $Z_{\max} = d \sin(6^\circ)$. This selection effect leads to a distance-dependent bias, where nearby sources probe only small heights above the plane, while more distant sources may appear at increasingly larger maximum heights. As a result, the observed vertical dispersion is not directly comparable across different regions of the Galaxy. To mitigate this effect, we define a dimensionless vertical stellar density dispersion,
\begin{equation}
\tilde{\sigma}_z = \frac{\sigma_z}{Z_\mathrm{max}},
\end{equation}
which normalizes the measured dispersion by the maximum accessible vertical height. In this form, $\tilde{\sigma}_z$ serves as a proxy for the intrinsic vertical scale height enabling a consistent comparison of the vertical structure across different Galactic environments.

Figure~\ref{fig:heatmap} shows the spatial distribution of $\tilde{\sigma}_z$ in the Galactic plane. The dispersion is computed in $(X,Y)$ bins, with the corresponding number of sources per bin shown for reference. Only bins with more than 35 objects have their dispersion computed. The map reveals a clear spatial variation, with the bulge region exhibiting higher, approximately constant values of $\tilde{\sigma}_z$ compared to the surrounding disk with $\tilde{\sigma}_z \approx 0.40$. Outside the bulge, the disk shows a uniform distribution with typical values near 0.30. If our ML-derived distances generally would place the sources significantly away from their actual distance, one would expect that the misplaced sources would artificially reduce (or enlarge) the vertical dispersion at locations closer (or further) from the Sun, most notably in the (therefore artificial) higher density regions. We do not see such a general gradient in distance from the Sun but instead, as mentioned, a variation between anticipated physical components of the MW --- the (local) disk and bulge.

To examine radial trends, we compute $\tilde{\sigma}_z$ as a function of Galactocentric radius ($R_{\rm GC}$), shown in Fig.~\ref{fig:dispersionplot}. The radial profile is constructed by first estimating $\tilde{\sigma}_z$ in spatial $(X,Y)$ bins and then averaging these cell-based values within concentric radial bins. The radial profile indicates that the dimensionless vertical dispersion is relatively high and nearly uniform within the bulge region, then declines toward larger radii, and finally remains approximately constant across the disk. This behavior is consistent with previous studies showing that the Galactic bulge is dynamically hotter and vertically thicker than the disk \citep{trapp2018sio}. In contrast, the disk exhibits a relatively thinner and more uniform vertical structure. The smooth transition observed in $\tilde{\sigma}_z$ is therefore consistent with the expected structural and dynamical differences between the bulge and disk components of the Galaxy at the anticipated Galactocentric distance.

To assess whether incompleteness in the most crowded inner-Galaxy region affects the estimated vertical structure, we repeated the analysis after excluding sources with $|l|<2^\circ$ and $|b|<1^\circ$. The resulting $\tilde{\sigma}_z$ profile remains consistent with the full-sample profile. Across the common radial bins, the median absolute fractional difference is only 0.34\%, with a maximum difference of 6.81\%. This indicates that removing the innermost crowded sightlines, a possible incompleteness in the plane, does not introduce an offset in the radial trend. This supports the robustness of the observed vertical-structure result and indicates that the resulting thickness is not artificially broadened by incompleteness in the innermost sightlines, further strengthening confidence in the derived distance estimates. As an additional check, we examined the source distribution in the \(X\)--\(Z\) plane. If severe near-plane incompleteness were driving the observed density structure, a clear deficit of sources near \(Z=0\) would be expected. Instead, the AKARI-selected sample remained concentrated toward the Galactic plane, with no strong depletion at \(Z=0\). This indicates that, although extinction and crowding affect the sample selection, severe near-plane incompleteness is not the dominant cause of the large-scale spatial structures discussed here.

To facilitate comparison with physically meaningful scale heights, we introduce a scaling factor $c$ and define an effective vertical dispersion,
\begin{equation}
\sigma_z = c \, \tilde{\sigma}_z.
\end{equation}
Here, $c$ may be chosen according to the reference vertical scale of interest. For example, adopting $c=1~\mathrm{kpc}$, corresponding to a characteristic thick-disk scale height \citep{tkachenko2025determining}, would map the observed dimensionless values to $\sigma_z \sim 300$--$400~\mathrm{pc}$. However, we leave $c$ as a general scaling factor, since the purpose of this relation is to provide a convenient conversion between the normalized dispersion and a physical vertical scale, rather than to impose a unique calibration.

The recovery of well-established Galactic structural trends using our distances provides an important validation of the adopted ML derived distance scale. Since the vertical dispersion depends directly on the inferred distances, any significant bias would distort the radial profile, either washing out the transition or introducing artificial gradients. While the normalization from $\tilde{\sigma}_z$ to $\sigma_z$ remains model-dependent, the relative trends are insensitive to the choice of scaling factor, and their consistency with the canonical Galactic structure supports the reliability of the ML-based distance estimates at the population level.

\subsection{Variability of AKARI AGB sample} \label{subsec:variab}

To confirm whether the considered AGB stars exhibit variable nature, we cross-match the cleaned \textit{AKARI}–2MASS catalog with the VSX catalog.
Additionally, this enables us to determine the kind of variability of the selected AGB sources, whether they are long-period Miras or semi-regular variables, which have significant differences in their mass, age, evolution and chemical characteristics. 

\subsubsection{Cross-matching with the VSX Catalog}

A positional matching radius of $5\arcsec$ was adopted, and in cases where multiple counterparts were identified, the reddest match 
was selected. This resulted in a total of 17,008 sources with reliable pulsation periods and variability classifications, with a majority (90\%) being classified as Mira or semiregular variables (SRVs), confirming that our color selection (Sect.\ \ref{sec:selection}) efficiently isolates late-type, mass-losing AGB stars. This final subset of confirmed variable AGB stars constitutes the working dataset used in our following variability analysis.

The primary source of uncertainty arises from potential contamination by C-rich AGB stars and YSOs, which can occupy similar mid-IR color space despite their distinct chemical and evolutionary properties. Using MSX/2MASS colors, \citet{lewis2020carbon} demonstrated an effective separation between O-rich and C-rich AGB stars, as well as a way to distinguish AGB stars from YSOs using the MSX D-E (15-21 $\mu$m) color. In the full BAaDE survey, the combined fraction of C-rich AGB stars and YSOs was estimated to be $8.3\%$ \citep{stroh2019bulge}. For the subset of sources exhibiting confirmed long-period variability, the likelihood of contamination by YSOs is expected to be minimal, as YSOs rarely display stable, coherent periodicities on timescales of several hundred days. Consequently, the dominant source of contamination in the variable AGB sample is expected to arise from C-rich AGB stars. Based on the MSX color analysis of \citet{lewis2020carbon}, the fraction of C-rich AGB stars in the BAaDE survey is estimated to be $\sim2.4\%$ which we consider acceptable as insignificant to be of influence in our analysis. Adopting a comparable fraction for our variable sample implies that approximately $410$ sources may be C-rich contaminants. The slightly contaminated sample of $\sim 17,000$ sources therefore, constitute a robust sample of O-rich variable AGB stars used in the subsequent analysis. 

We further performed a cross-match with the near-IR Mira catalog of \citet{sanders2022mira} to compare our spatial distribution with theirs. Using our combined AKARI and MSX sample, we obtained only 12 crossmatches. The limited overlap is therefore primarily due to the lack of AKARI measurements. This is consistent with the AKARI coverage shown by their work, where majority of their sources do not have AKARI counterparts (Fig.\ 14 in \citet{sanders2022mira}). For the few matched sources, the median distance we obtain is approximately 8.5 kpc, close to the Galactic-center distance and consistent with an nuclear stellar disk association, as expected for many of the near-IR Mira sources in their sample. However, because of the small number of matched sources, this catalog does not provide a statistically meaningful comparison of the spatial distribution of our sample and we focus on the VSX comparison as the primary variability-based check. The limited overlap with the \citet{sanders2022mira} sample indicates that our catalog is incomplete in highly extincted and severely crowded regions, including the nuclear stellar disk, because of the sensitivity and angular-resolution limitations of 2MASS and AKARI. Consequently, the present analysis does not adequately probe the nuclear stellar disk or similarly challenging inner-Galaxy environments.

\begin{figure*}[t]
    \centering
    \includegraphics[trim=0.0cm 0.0cm 0.0cm 0.0cm,clip,width=\textwidth]{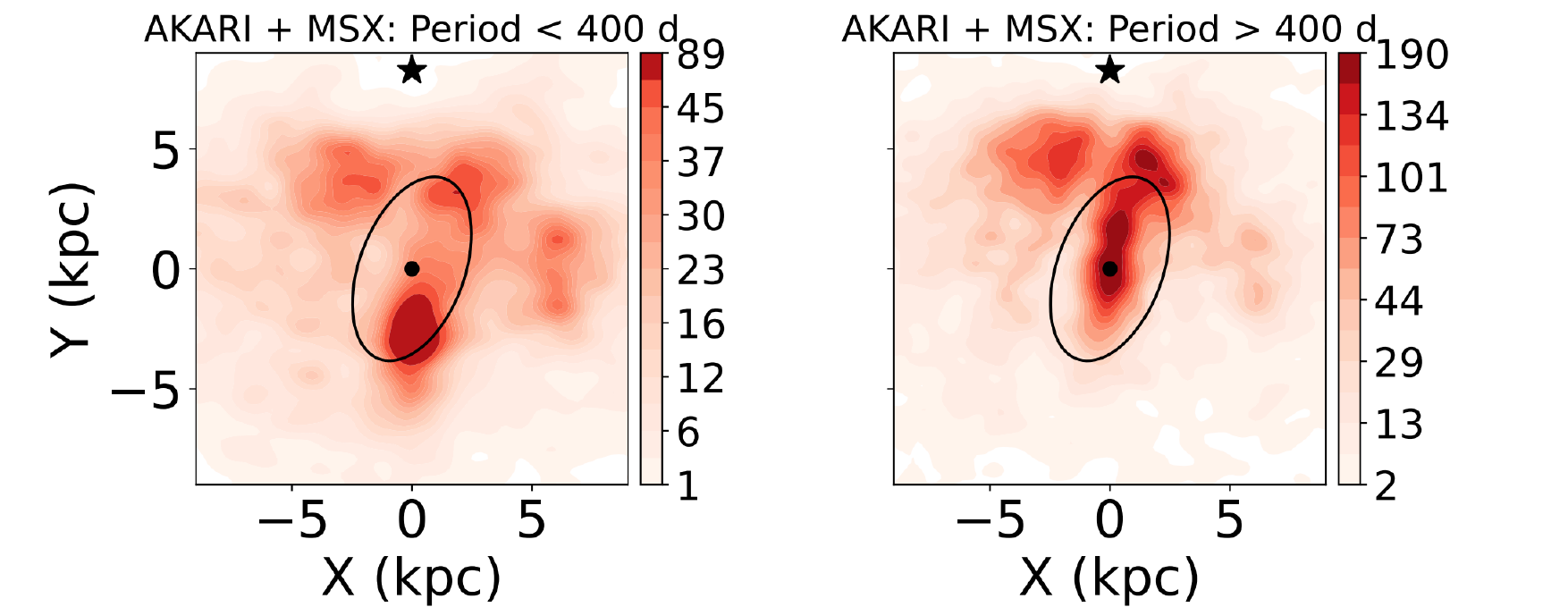}
    \caption{Spatial distributions of Mira variables in the Galactic bulge as a function of pulsation period. The left panel shows the distribution of short-period Miras ($P<400$ d), while the right panel shows the distribution of long-period Miras ($P>400$ d), using ML-based distances for the combined \textit{AKARI}+MSX sample. The MSX distances are taken from \citet{Bhattacharya2024}, with the additional requirement MSX-A $<4.5$. Coordinates are shown in the Galactic plane, with the Galactic center located at $(X,Y)=(0,0)$; the black star marks the Solar position. The black ellipse represents the adopted bulge/bar boundary, inclined by $20^\circ$ following \citet{iwanek2023three}. The color shading indicates the relative surface density of sources in units of $N\,\mathrm{kpc}^{-2}$, where $N$ is the number of sources.}
    \label{fig:barstruc}
\end{figure*} 

\begin{figure*}[t]
    \centering
    \includegraphics[trim=0.0cm 0.0cm 0.0cm 0.0cm,clip,width=\textwidth]{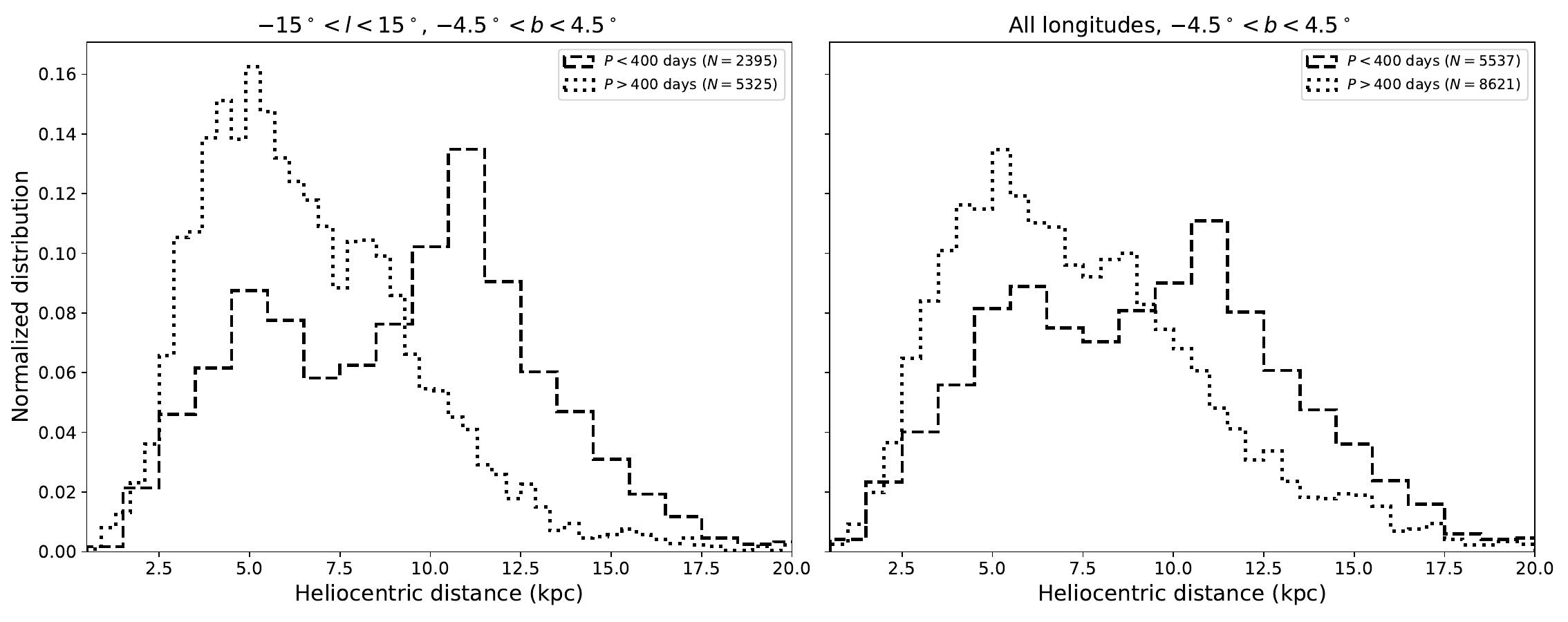}
    \caption{Normalized heliocentric-distance distributions of Mira variables separated by pulsation period. The left panel shows sources within $-15^\circ<l<15^\circ$, while the right panel includes all Galactic longitudes; both panels are restricted to $-4.5^\circ<b<4.5^\circ$ following \citet{catchpole2016age}. Dashed histograms represent short-period Miras ($P<400$ d), and dotted histograms represent long-period Miras ($P>400$ d).} 
  
    \label{fig:periodhist}
\end{figure*}

\begin{figure*}[t]
    \centering
    \includegraphics[trim=0.0cm 0.0cm 0.0cm 0.0cm,clip,width=\textwidth]{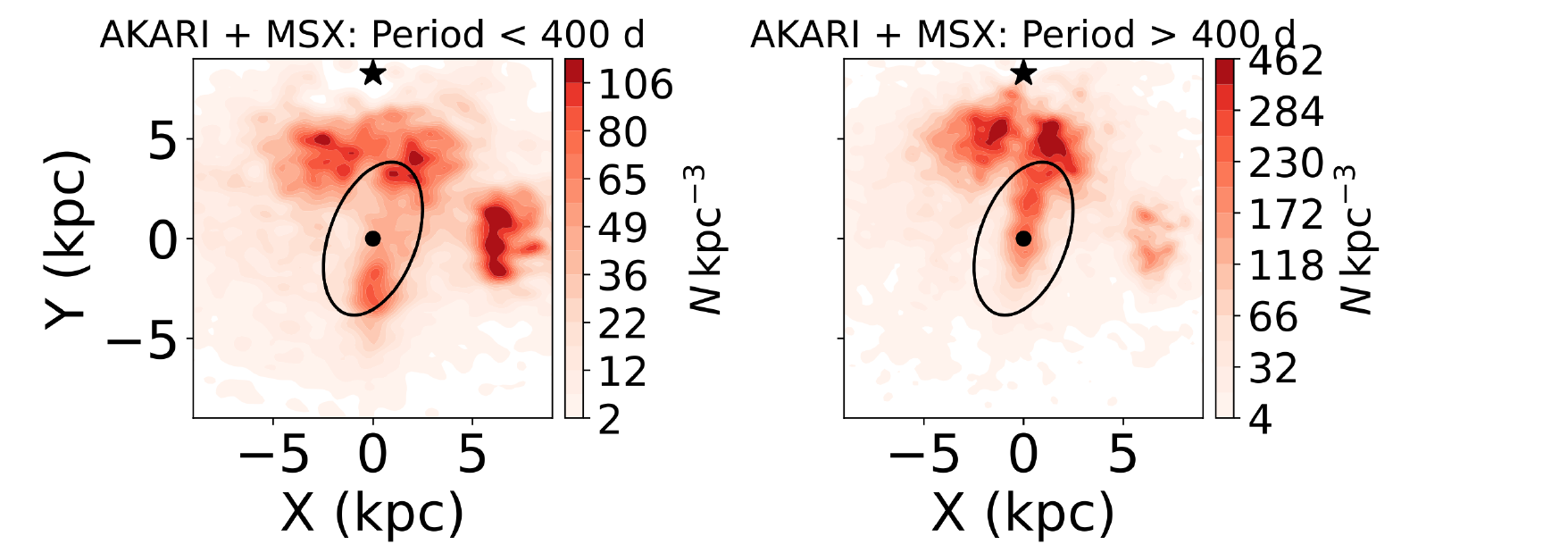}
    \caption{Volume-corrected spatial distributions of Mira variables in the Galactic bulge as a function of pulsation period. The left panel shows the distribution of short-period Miras ($P<400$ d), while the right panel shows the distribution of long-period Miras ($P>400$ d), using ML-based distances for the combined \textit{AKARI}+MSX sample. The MSX distances are taken from \citet{Bhattacharya2024}, with the additional requirement MSX-A $<4.5$. Coordinates are shown in the Galactic plane, with the Galactic center located at $(X,Y)=(0,0)$; the black star marks the Solar position. The black ellipse represents the adopted bulge/bar boundary, inclined by $20^\circ$ following \citet{iwanek2023three}. Color shading indicates the volume-corrected source density in units of $N\,\mathrm{kpc}^{-3}$, where $N$ is the number of sources. After applying the volume correction, the prominent far-side clump seen in the projected density map is substantially reduced, and the corrected distribution appears more uniform, with the far-side enhancement becoming comparable to the near-side structure.}
    \label{fig:barstrucvol}
\end{figure*}

\subsubsection{Discerning the bar structure with longer period Miras}

Figure~\ref{fig:barstruc} compares the face-on spatial distributions of Mira variables in the inner Galaxy as a function of pulsation period. For the purpose of discerning the bar structure, we restrict the analysis to Mira variables, as their pulsation periods are more reliably determined than those of SRVs. It is found that the short-period, older Miras ($P \leq 400$~d) exhibit a relatively smooth distribution within the bulge, with little evidence for a coherent elongated structure. In contrast, long-period Miras ($P > 400$~d) show a markedly stronger central concentration and a pronounced elongation aligned with the Galactic bar. This clear morphological distinction highlights the sensitivity of Mira populations to stellar age and evolutionary state. The barred morphology traced by the long-period population becomes even more apparent when incorporating Mira variables from the MSX-selected sample with previously determined SED-based distances \citep{Bhattacharya2024} and pulsation periods obtained from the VSX catalog, which provide a more complete sampling of AGB stars across the Milky Way, particularly in the solar neighborhood.

In Fig.~\ref{fig:barstruc}, we restrict the MSX sample to sources with MSX-A $<4.5$ to minimize contamination from deeper, non-uniform scans \citep{Egan_MSX1999}. This selection homogenizes longitude coverage across the inner Galaxy and improves the recovery of large-scale spatial structure, especially the Galactic bar, without altering the observed period-dependent trends. The persistence and strengthening of the bar signature in the combined \textit{AKARI}+MSX sample demonstrate that the observed elongation is not driven by selection effects in the \textit{AKARI} data.

The bulge ellipse shown in Fig.~\ref{fig:barstruc} is drawn with a position angle of $20^{\circ}$, following \citet{iwanek2023three}, who derived this inclination using a large, independent sample of Mira variables. Although we adopt a slightly larger bar angle of $27^{\circ}$ for the broader AGB candidate sample, we use the Mira-specific inclination here to maintain consistency with the tracer population. This difference does not affect our conclusions, as the qualitative separation between short- and long-period Mira distributions remains robust to reasonable variations in the assumed bar geometry. The distinct morphological behavior of the two Mira populations is consistent with the results of \citet{catchpole2016age}, who demonstrated that the bulge structure traced by Mira variables depends strongly on stellar age. In their analysis, longer-period Miras (younger) traced a barred morphology, whereas shorter-period Miras (older) were more consistent with a spheroidal distribution. 

From an evolutionary perspective, O-rich Mira variables at (sub)solar metallicity are expected to evolve into C-rich AGB stars through repeated TDU episodes. However, stellar evolution models predict that in more massive AGB stars ($\mathrm{M} \gtrsim 3.5,\mathrm{M_\odot}$), this transition can be delayed or suppressed due to Hot Bottom Burning (HBB), which destroys carbon in the stellar envelope \citep{jimenez2015study}. As a result, such stars can remain O-rich throughout much of their AGB evolution. Long-period Miras ($P>400$~d), often associated with HBB candidates \citep{Whitelock_hbb_2003}, are therefore expected to be relatively more massive and younger. 

The strong concentration of long-period, O-rich Miras along the Galactic bar is thus consistent with a scenario in which the bar hosts a substantial population of relatively young, massive, and metal-rich AGB stars that preferentially trace its elongated structure. This interpretation aligns with independent evidence linking the Galactic bar to metal-rich stellar populations: for example, old and metal-poor stars such as RR Lyrae or Type II Cepheids do not exhibit a barred distribution, whereas younger and more metal-rich red clump stars clearly trace the bar \citep{bhardwaj2017galactic, braga2018structure, cabrera2007tracing}. 

We also examined the Mira distance distribution as a function of pulsation period in comparison with \citet{catchpole2016age}. We divided the sample into $\Delta\log P=0.1$ intervals over $2.1<\log P\leq2.7$ and calculated the mean heliocentric distance in each bin for both the full longitude range and the restricted region $-15^\circ<l<15^\circ$, while retaining $-4.5^\circ<b<4.5^\circ$. The uncertainty on each mean was obtained by propagating the adopted 35.5\% uncertainty of the individual distance estimates. These results are listed in Table~\ref{tab:compcatchpole} and compared with the corresponding $R_0$ values reported by \citet{catchpole2016age}, which were obtained from least-squares fits of distance modulus as a function of Galactic longitude. For the combined short-period interval, $2.1<\log P\leq2.6$, we use all Miras over the full distance span and obtain mean distances of $9.23\pm0.05,\mathrm{kpc}$ over all longitudes and $9.17\pm0.07,\mathrm{kpc}$ within $-15^\circ<l<15^\circ$, in close agreement with the corresponding value of $R_0=8.99\pm0.04,\mathrm{kpc}$ from \citet{catchpole2016age}. The largest difference occurs in the shortest-period interval, $2.1<\log P\leq2.2$, where our all-longitude mean is $10.39\pm0.66\,\mathrm{kpc}$ compared with $9.09\pm0.22\,\mathrm{kpc}$ from \citet{catchpole2016age}. This bin contains only 38 sources and consequently has the largest uncertainty. In the better-populated bins with $2.3<\log P\leq2.6$, the differences are only $\sim0.1$--$0.3\,\mathrm{kpc}$. For the longest-period interval, $2.6<\log P\leq2.7$, our all-longitude mean distance of $8.12\pm0.04\,\mathrm{kpc}$ is also close to the value of $8.33\pm0.08\,\mathrm{kpc}$ reported by \citet{catchpole2016age}, although the restricted $-15^\circ<l<15^\circ$ sample has a smaller mean distance of $7.37\pm0.05\,\mathrm{kpc}$.

The heliocentric-distance histograms of the two broad period groups are shown in Fig.~\ref{fig:periodhist}. The two samples have broadly similar near-side distributions but differ more clearly around the Galactic-center and far-side distance ranges. Within the more bulge-focused region, $-15^\circ<l<15^\circ$, both samples show a peak near $5$--$6\,\mathrm{kpc}$. The long-period sample ($P>400$ d) contains a stronger concentration near the GC distance, consistent with the centrally concentrated and elongated structure seen in Fig. \ref{fig:barstruc}. By contrast, the short-period sample ($P<400$ d) shows near- and far-side peaks of comparable strength, with the far-side enhancement near $10$--$11\,\mathrm{kpc}$ being more pronounced. The location of the far-side peak does not necessarily coincide with the arithmetic mean because the short-period distribution also contains substantial foreground and intermediate-distance components.We also find broad agreement with \citet{catchpole2016age} across the period bins, suggesting that the far-side enhancement is unlikely to arise solely from a systematic overestimation of the distances to short-period Miras, although uncertainties in the distance estimates, sample-selection effects, and projection effects may still enhance its apparent prominence, the extent of these contributions remains uncertain at present.

Importantly, we emphasize that pulsation period is not included as an input feature in our ML distance model, and the distances are estimated before the sample is divided into short- and long-period groups. The model therefore has no explicit mechanism to assign systematically larger distances to ($P<400$) d Miras and smaller distances to ($P>400$) d Miras on the basis of period. The difference between the two distance distributions emerges only after the independently estimated distances are separated by period. A systematic bias that is independent of period or other stellar properties would be expected to affect sources across a broad period range, rather than selectively placing short-period Miras on the far side while assigning shorter distances to long-period Miras. The additional far-side concentration seen in the short-period sample is substantially reduced after applying the distance-dependent vertical projection correction shown in Fig.~\ref{fig:barstrucvol}, indicating that projection effects contribute significantly to its apparent prominence. A similar far-side enhancement is visible in the AKARI-based O-rich AGB distribution of \citet{ishihara2011galactic} (their Fig.~8c), where the assumed GC distance is $8.5,\mathrm{kpc}$ and the overdensity extends several kpc beyond it along the line of sight. Since those distances were obtained using a different AKARI-based photometric method involving the AKARI 9~$\mu$m flux and $[K]-[9]$ color, this agreement suggests that the feature is unlikely to be produced solely by a systematic bias specific to our distance-estimation method.

Nevertheless, the bimodal distance distribution shown in Fig.~\ref{fig:periodhist} for the short-period Miras may indicate the presence of systematic uncertainties in the distance estimates, although selection effects associated with the AKARI sample and projection effects may also contribute to its observed shape. In particular, period-correlated changes in Mira SEDs, including variations in color and circumstellar-dust emission, could introduce indirect systematic effects even though period is not explicitly used in the distance-estimation model. We therefore do not interpret this far-side enhancement as an additional physical result at present, and its detailed origin requires further investigation.

\begin{table*}[t]
\centering
\caption{Mean distances of the Mira sample in different period ranges for sources with $-4.5^\circ<b<4.5^\circ$. The quoted uncertainties for our sample are obtained by propagating the adopted 35.5\% uncertainty of the individual distance estimates. For comparison, the values from \citet{catchpole2016age} are their fitted $R_0$ estimates for the all-longitude sample in the same latitude and period ranges.}
\label{tab:period_distance_comparison}
\begin{tabular}{lcccccc}
\hline
\multirow{2}{*}{Log period range}
& \multicolumn{2}{c}{All longitudes}
& \multicolumn{2}{c}{$-15^\circ<l<15^\circ$}
& \multicolumn{2}{c}{\citet{catchpole2016age}} \\
\cline{2-7}
& $\langle d\rangle$ (kpc) & $N$
& $\langle d\rangle$ (kpc) & $N$
& $R_{0,\rm Catchpole}$ (kpc) & $N_{\rm Catchpole}$ \\
\hline
\multicolumn{7}{c}{$-4.5^\circ<b<4.5^\circ$} \\
\hline
$2.1<\log P\leq2.2$
& $10.39\pm0.66$ & 38
& $9.71\pm0.85$ & 21
& $9.09\pm0.22$ & 142 \\

$2.2<\log P\leq2.3$
& $9.99\pm0.30$ & 168
& $9.57\pm0.41$ & 85
& $8.98\pm0.14$ & 384 \\

$2.3<\log P\leq2.4$
& $9.27\pm0.20$ & 326
& $8.98\pm0.29$ & 163
& $8.95\pm0.08$ & 816 \\

$2.4<\log P\leq2.5$
& $9.32\pm0.13$ & 795
& $9.47\pm0.21$ & 311
& $9.19\pm0.07$ & 1579 \\

$2.5<\log P\leq2.6$
& $9.16\pm0.06$ & 4063
& $9.11\pm0.08$ & 1755
& $8.83\pm0.04$ & 1962 \\

$2.1<\log P\leq2.6$
& $9.23\pm0.05$ & 5390
& $9.17\pm0.07$ & 2335
& $8.99\pm0.04$ & 4883 \\

$2.6<\log P\leq2.7$
& $8.12\pm0.04$ & 6310
& $7.37\pm0.05$ & 3492
& $8.33\pm0.08$ & 1182 \\
\hline
\label{tab:compcatchpole}
\end{tabular}
\end{table*}

\section{Summary and Conclusions}\label{sec:summary_conclusions}

In this work, we have developed and applied a supervised machine-learning framework to estimate statistical distances to dust-obscured, O-rich AGB stars using multi-band IR photometry. By training an XGBoost regression model on the BAaDE AGB sample with previously derived SED-based distances, we extend reasonable distance estimates ($\pm 36\%$) to a much larger population of evolved stars selected from the \textit{AKARI} mid-IR survey. The main conclusions of this study are summarized as follows:

\begin{enumerate}
    \item The optimized XGBoost distance regressor  maps near- and mid-IR photometry to stellar distance, achieving strong and stable performance with $R^{2}\approx0.98$ on an independent test set. This demonstrates that multi-band IR photometry alone contains sufficient information to recover statistical distances for heavily obscured AGB stars irrespective of their variable nature.
    
    \item Applying the trained model to the \textit{AKARI}/IRC-color selected O-rich AGB stars in the  \textit{AKARI} catalog yields distance estimates for more than 36,000 additional sources, substantially enhancing the available distance information for evolved stars in the inner MW. This represents a significant improvement over previous VLBI/Gaia/PLR/MSX-based samples, extending the accessible volume several kpc farther into the Galactic plane and bulge.
    
    \item We validate the inferred distance scale through comparisons with independent P-L relations and published distances for Galactic Mira variables. In particular, the model-predicted distances show improved agreement with mid-IR P-L relations calibrated on Galactic maser-bearing Miras and are consistent, within uncertainties, with the small subsample cross-matched with \citet{urago20203d}. Despite being statistical with large relative errors for individual sources, these comparisons show that ensemble distances are physically meaningful.

     \item  We investigate the vertical structure of the inner Galaxy and find that the spatial and radial distributions exhibit a clear trend: the vertical dispersion is relatively high and nearly uniform within the bulge, decreases with increasing radius, and becomes approximately constant across the disk. The lack of artificial gradients with distance from the Sun, combined with the distinct transition between bulge and disk, suggests that the inferred distances reliably capture the underlying Galactic structure. These results are consistent with a vertically thicker, dynamically hotter bulge and a thinner, more uniform disk.
    
    \item Using the expanded distance catalog, we also investigate the spatial distribution of Mira variables in the Galactic bulge. Long-period Miras preferentially trace an elongated barred morphology. This behavior is consistent with previous studies suggesting that younger, more massive AGB populations are more strongly associated with the Galactic bulge. This consistency across independent tracers reinforces both the physical interpretation of the observed trends and the reliability of our ML-based distance estimates.

    \item Short-period Mira variables ($P \leq 400$~d) show a relatively smooth and spheroidal distribution in the bulge, with little evidence of a bar-like structure. This is consistent with an older population that is more dynamically mixed, in contrast to long-period Miras that clearly trace the Galactic bar. However, the broad and potentially bimodal distance distribution of the short-period Miras may also be affected by projection, sample-selection, or systematic uncertainties in the distance estimates, and its detailed origin will be investigated in future work.

\end{enumerate}

This study demonstrates that ML-based distance estimation from IR photometry can scale up and provide collective distance information to large samples of dust-obscured AGB stars. The distance catalog recovers large-scale Galactic features like the bar and spiral arms and establishes long-period Mira variables as vital tracers of younger stellar populations in the bulge and disk. In the future, combining statistical distances with BAaDE maser kinematics, enhanced extinction modeling, and upcoming IR time-domain surveys like JWST, Roman, and Rubin/LSST will improve chemo-dynamical constraints on the formation and evolution of the inner MW.

\section{Acknowledgements}

Support for this work was provided by the National Science Foundation (NSF) through
the Grote Reber Fellowship Program administered by Associated Universities, Inc./National Radio
Astronomy Observatory (AUI/NRAO). The National Radio Astronomy Observatory is a facility of the U.S.\ National
Science Foundation operated under cooperative agreement by Associated Universities, Inc. SK and AB acknowledge IUCAA for the financial support.


\bibliography{References}{}
\bibliographystyle{aasjournalv7}



\end{document}